\documentclass[twocolumns]{aa}
\usepackage{graphicx}
\usepackage{amssymb}
\usepackage{amsmath}
\usepackage{rotate}
\usepackage{cancel}

\begin{document}

\title{On the fine structure of sunspot penumbrae\\
III. The vertical extension of penumbral filaments}
\author{Borrero, J.M.\inst{1,}\inst{2} \and Solanki, S.K.\inst{1}
\and Lagg, A.\inst{1} \and Socas-Navarro, H.\inst{2} \and Lites, B.\inst{2}}
\institute{Max--Planck Institut f\"ur Sonnensystemforschung, 37191,
Katlenburg-Lindau, Germany \and High Altitude Observatory, 3450 Mitchell Lane, 
Boulder, 80301 Colorado USA.}

\abstract{In this paper we study the fine structure of the penumbra as inferred
from the uncombed model (flux tube embedded in a magnetic surrounding) when applied 
to penumbral spectropolarimetric data from
the neutral iron lines at 6300 \AA. The inversion infers very
similar radial dependences in the physical quantities (LOS velocity, magnetic
field strength etc) as those obtained from the inversion of the Fe I
1.56 $\mu$m lines. In addition, the large Stokes $V$ area
asymmetry exhibited by the visible lines helps to constrain the size of the
penumbral flux tubes. As we demonstrate here, the
uncombed model is able to reproduce the area asymmetry with striking
accuracy, returning flux tubes as thick as 100-300 kilometers  in the vertical
direction, in good agreement with previous investigations.
}
\authorrunning{{\bf Borrero et al.}}
\titlerunning{{\bf vertical extension of the penumbral filaments}}
\maketitle

\newcommand{\ve}[1]{\rm{\bf {#1}}}
\newcommand{\syn}{_{\rm syn}}
\newcommand{\obs}{_{\rm obs}}
\newcommand{\wl}{(\lambda)}
\newcommand{\ekis}{${\rm{\bf X}}$}
\newcommand{\ekiss}{{\rm{\bf X}}}

\section{Introduction}%

There is already considerable evidence pointing towards a magnetic
structure of the sunspot penumbra in the form of uncombed magnetic 
fields: Mart{\'\i}nez Pillet (2000, 2001), Schlichenmaier et al. (2002),
M\"uller et al. (2002), Mathew et al. (2003), Bellot Rubio et al. (2003, 2004), 
Borrero et al. (2004, 2005). The {\it uncombed} magnetic field was first 
employed by Solanki \& Montavon (1993) to qualitatively reproduce 
the net circular polarization (NCP) observed in 
the sunspot penumbra. The uncombed model, as proposed by these 
authors, consisted of a three layered atmosphere in which the central layer would 
correspond to a {\it horizontal flux tube} carrying the Evershed flow. 
As the net circular polarization can only be produced by gradients along the line
of sight in the physical parameters relevant to the line formation (magnetic field vector,
temperature, velocity, etc.), the uncombed model was originally conceived as a model to explain the
vertical inhomogeneities of the sunspot penumbra as observed at moderate spatial
resolution ($\sim$1 arc sec).

This model was later generalized by Mart\' \i nez Pillet (2000), who realized that
 the horizontal extension of the flux tube could be smaller than the size of 
the resolution element. To account for this possibility, he added a 
second component, {\it surrounding}, that would lie next to the three
layered atmosphere from Solanki \& Montavon and whose properties would 
be the same as those of the uppermost and lowermost layers. With this, Mart{\'\i}nez
Pillet attempted to model the horizontal inhomogeneities in the penumbral structure.
Besides he was able to provide a qualitatively explanation for the 
Center-to-Limb variation of the net circular polarization in sunspot penumbrae
and found that either the {\it surrounding} atmosphere would have a non vanishing
velocity, or the magnetic flux tubes had to be pointing downwards (i.e.: 
returning to the solar surface) in the outer penumbra (in agreement with 
previous findings from Westendorp Plaza et al. 1997).

Mathew et al. (2003) qualitatively demonstrated that the uncombed model 
was also able to reproduce the different results obtained from the
 separate inversion of the Fe I lines at 6300 \AA~ (Westendorp Plaza 
et al. 2001a, 2001b) and 1.56 $\mu$m when a geometrical model consisting 
of one atmosphere with depth-dependent physical parameters 
was used. More recently, Borrero et al. (2004; hereafter Paper I), demonstrated 
that results obtained from the inversion of penumbral spectropolarimetric 
observations in the 1.56 $\mu$m lines using one atmospheric component with 
gradients in optical depth and two components without gradients were perfectly 
compatible with each other, both pointing towards a picture for the penumbral 
fine structure based on the uncombed model. These authors ascribed their
results to the inability of the Fe I lines at 1.56 $\mu$m to distinguish
between the vertical and the horizontal inhomogeneities in a magnetic
atmosphere. Borrero et al. (2005; hereafter Paper II) then
applied the uncombed model to full spectropolarimetric observations of the infrared Fe I 
lines at 1.56 $\mu$m. As this set of lines is not sensitive enough to gradients along
the line of sight to exhibit a significant amount of net circular polarization,
their inversions returned flux tubes whose vertical extensions were
much larger than the formation height range of the employed spectral lines.
Therefore the uncombed model would be almost equivalent to a geometrical model using
two components without gradients along the line of sight (as in Paper I; see also
Bellot Rubio et al. 2003,2004).

Very recently, the uncombed model has been confronted against 
polarimetric observations at high spatial resolution (0.2 arc sec; 
Langhans et al. 2005; Bellot Rubio et al. 2005). These authors were able, 
for the first time, to resolve in individual polarization signals 
the horizontal magnetic structure of the sunspot penumbra in a pattern
of horizontal flux tubes that carry the Evershed flow and a less 
inclined magnetic surrounding. This demonstrates, in agreement
with previous works (S\"utterlin 2001; Scharmer et al 2002; 
S\"utterlin et al. 2004; Rouppe van der Voort et al. 2004)
that the azimuthal width (i.e. horizontal extension) of the penumbral filaments 
is of the order of 0.2 arc sec or equivalently about 150 kilometers. 

Yet, the vertical extension of the penumbral flux tubes is not properly
known. In this work we will follow the suggestions made in Papers I and II,
where we already pointed out that the visible Fe I lines at 6300 \AA~ are not 
only much more sensitive to line of sight gradients than their 
infrared counterparts, but are also formed over a much larger optical
 depth range, and therefore they might help to constrain the vertical extension 
of the penumbral flux tubes.

\section{Observations}%

The active region NOAA 8545 was observed on May 21st, 1999 at a heliocentric 
angle of $\mu=\cos\theta=0.79$ using the ASP (Advance Stokes Polarimeter; 
Elmore et al. 1992) instrument. The recorded spectral range of
3.2 \AA~width is sampled every 12.8 m\AA. It contains the full Stokes
vector of four Zeeman-sensitive spectral lines: Fe I (g$_{\rm eff}$=1.67) 6301.5 \AA~, Fe I 
(g=2.5) 6302.5 \AA~, Fe I (g$_{\rm eff}$=1.5) 6303.4 \AA~ and Ti I (g$_{\rm eff}$=0.92)
6303.7 \AA. Note that the first two neutral iron lines are blended by two telluric
O$_2$ lines in the far red wing.

The data reduction was performed following the usual procedure (see Skumanich et al. 1997). 
As always, special care was taken in the wavelength calibration, for which 
we first assumed that the average umbral profile of 6301.5 and 6302.5 \AA~ is at rest 
(i.e. Stokes I center of gravity corresponds to the central laboratory
wavelengths) and later a minor correction was done at each pixel using
the telluric lines. The laboratory wavelength for the Ti I line,  as well as
the oscillator strengths of Fe I 6303.4 \AA~ and Ti I 6303.7 \AA~ are
inaccurate, and therefore we recalculated them using the two component model
for the quiet Sun obtained by Borrero \& Bellot Rubio (2002) using the procedure described in
Borrero et al. (2003). The relevant atomic parameters for the four lines are
presented in Table 1. Although these lines are less sensitive to the magnetic field than their
infrared counterparts (see Papers I and II), they present a number of
advantages that make them very suitable for our purposes.
The visible lines provide a better height coverage as they are formed 
over a wider range of optical depth layers in the solar photosphere: 
$\log\tau_5 \in [-4,0]$. In addition, as already pointed out in 
Papers I and II, the visible Fe I lines are far
more sensitive to gradients along the line of sight in the physical
quantities. Finally, the visible lines are more affected 
by changes in the temperature stratification than the Fe I lines at 1.56
$\mu$m. This is mainly of advantage in the outer penumbra, where the 
OH lines that blend the Fe I 15652 \AA~ line are too weak to be used.

\begin{table}
\caption[]{Atomic parameters of the observed lines. $\lambda_0$ 
represents the laboratory central wavelength, $\chi_{\rm l}$ the
excitation potential of the lower energy level, and $\log gf$ the 
logarithm of the oscillator strength times the multiplicity of
the level. The parameters $\alpha$ and $\sigma$ (in units of 
Bohr's radius, $a_0$) are used to calculate the broadening of
the lines by collisions with neutral hydrogen atoms as resulting 
from the ABO theory (Barklem \& O'Mara 1997; Barklem et al. 1998). 
The seventh column gives the effective Land\'e factor of the transition, $g_{\rm eff}$.}
\begin{center}
\tabcolsep .6em
\begin{tabular}{lcccccc}
 \hline
Species & $\lambda_{0}$ & $\chi_{l}$ & log~gf & $\alpha$ &
$\sigma$ & $g_{\rm eff}$\\
& (\AA~) & (eV) & (dex) & & ($a_{0}^{2}$)\\\hline
\hline
Fe I & 6301.5012 &     3.654 &    $-$0.718  &  0.243  & 832 & 1.67\\           
Fe I & 6302.4916 &     3.686 &    $-$1.235  &  0.240  & 847 & 2.50\\
Fe I & 6303.4600 &     4.320 &    $-$2.550  &  0.276  & 712 & 1.50\\           
Ti I & 6303.7560 &     1.443 &    $-$1.611  &  0.236  & 357 & 0.92\\
\hline
\end{tabular}
\end{center}
\end{table}

Figure 1 shows maps of the observed active region as seen in
the continuum intensity (top panel), total unsigned circular polarization
(middle panel) and the normalized NCP or area asymmetry (bottom panel).
In all these three plots the direction of the center of the Solar disk is
indicated by the white arrow. The dashed white line points also towards the
center of the Solar disk passing through the center of the umbra (defined as
the darkest point seen in the continuum image) and therefore indicating
the so-called line of symmetry of the sunspot. The square box indicates the region on
the limb side of the penumbra that has been selected for our study.

\begin{figure}
\begin{center}
\includegraphics[width=8cm]{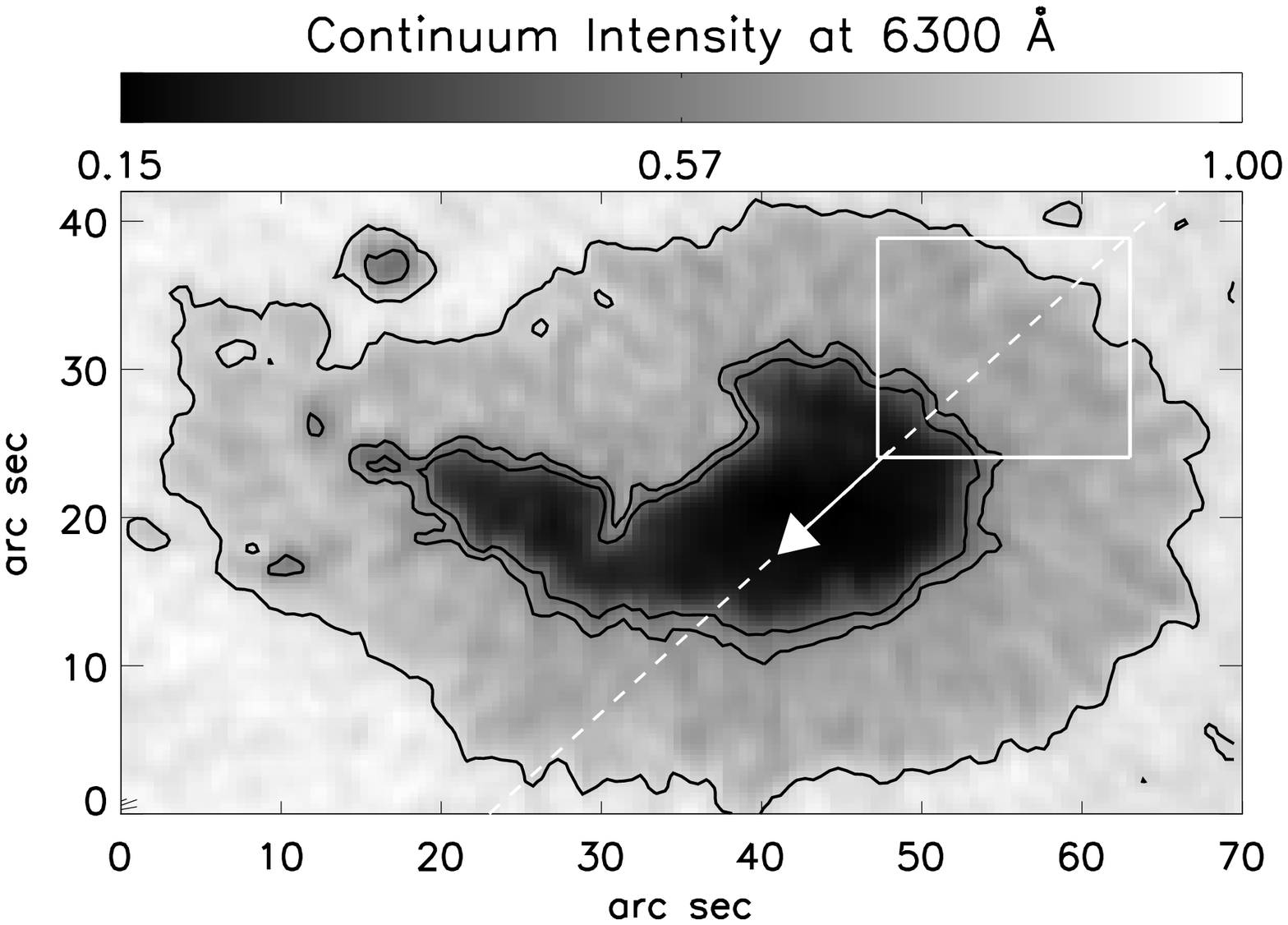}
\includegraphics[width=8cm]{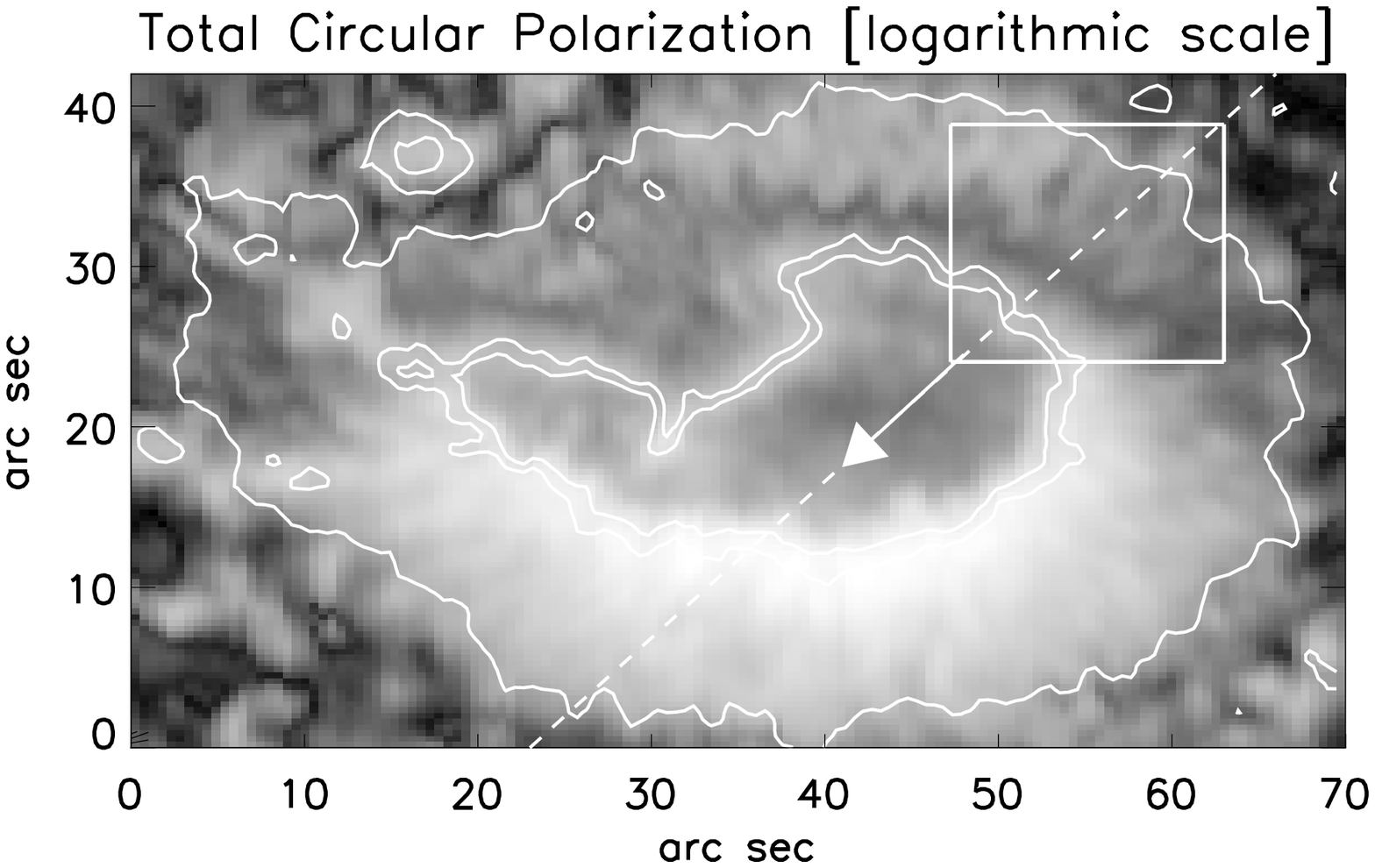}
\includegraphics[width=8cm]{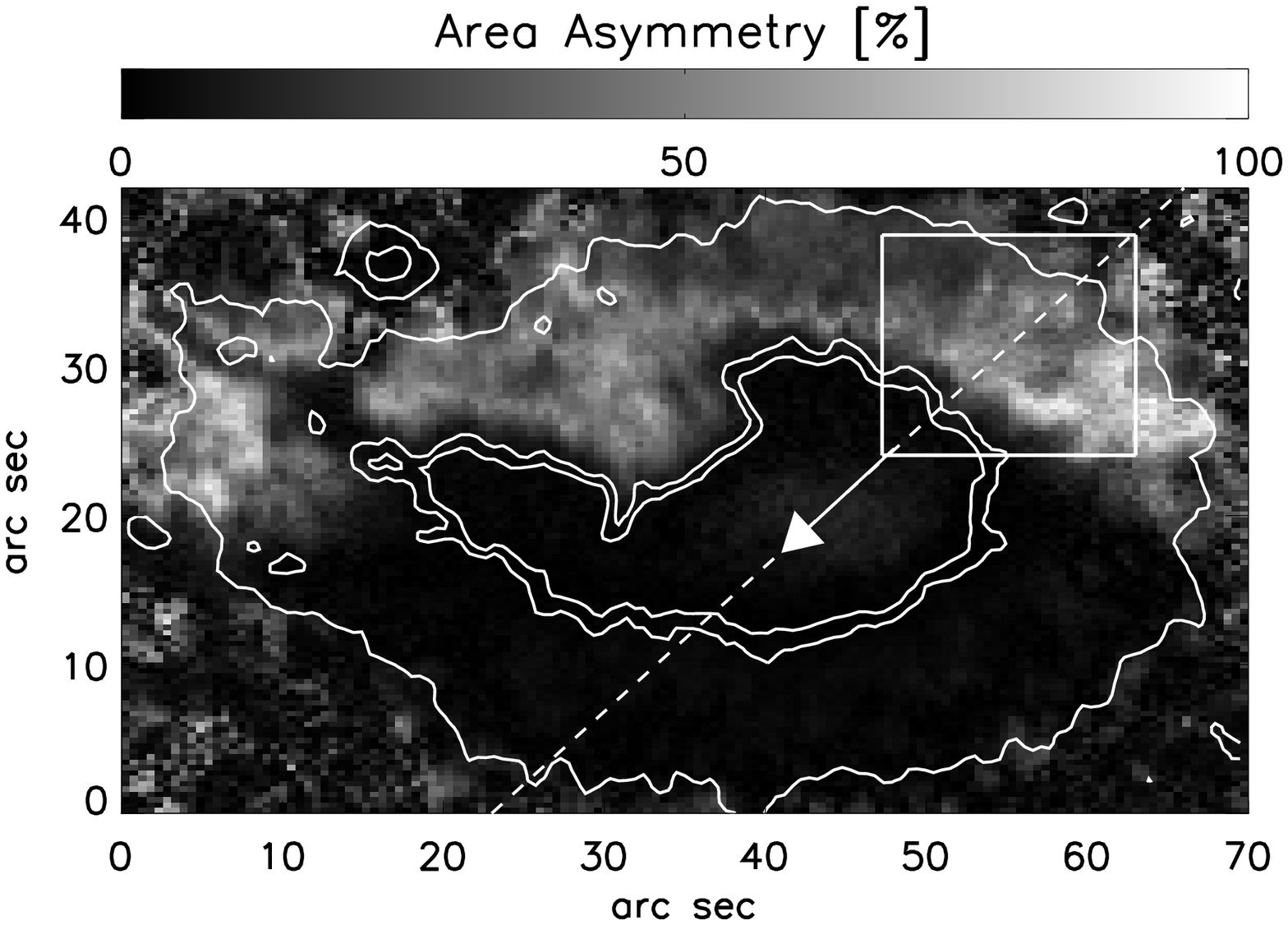}
\end{center}
\caption{Continuum intensity (top panel), total circular
polarization map (middle panel) and area asymmetry maps (bottom panel) of
NOAA 8545. The contours correspond to the levels 0.45, 0.6 and 0.85 in units of the quiet sun
intensity I$_C$. The outermost boundary defines the sunspot radius $r=R$.
The arrow points towards the direction of the disk
center passing through the darkest umbral region (this defines the line
of symmetry). The selected region for the inversion is indicated by the white 
square box on the limbward side of the penumbra, centered on the line of symmetry 
of the sunspot, where the area asymmetry reaches the largest values.}
\end{figure}

\section{Uncombed model and inversion strategy}%

Our inversion strategy considers a geometrical
model, defined by a set of physical parameters  
\ekis~ (magnetic field vector, temperature, velocity) 
that enter the Radiative Transfer Equation and are used to produce synthetic Stokes profiles.
By comparing the synthetic profiles and the observed ones,
 the original set of parameters is iterated in some 
fashion until convergence is achieved. The geometrical 
model we have selected as representative of the penumbral fine 
structure is the same as in Paper II, id est: a modification
of the uncombed model of Solanki \& Montavon (1993).
A detailed description of the implementation of the uncombed model
into an existing inversion code and its peculiarities can be found 
in Paper II (Sect.~2). Here we simply discuss its main 
features for the sake of completeness.

In order to produce synthetic Stokes profiles
under the uncombed model, the Radiative Transfer Equation 
is solved along two different ray paths (see Fig.~2), one of them piercing 
the flux tube cross section (dashed line) and another passing 
through the atmosphere that surrounds the flux tube (dot-dashed line). 
In this simplification the flux tube has a rectangular cross section.
The synthetic Stokes vector, $\ve{S}\syn\wl$, to be compared with the observed
ones is obtained according to:

\begin{figure}
\begin{center}
\includegraphics[width=8cm]{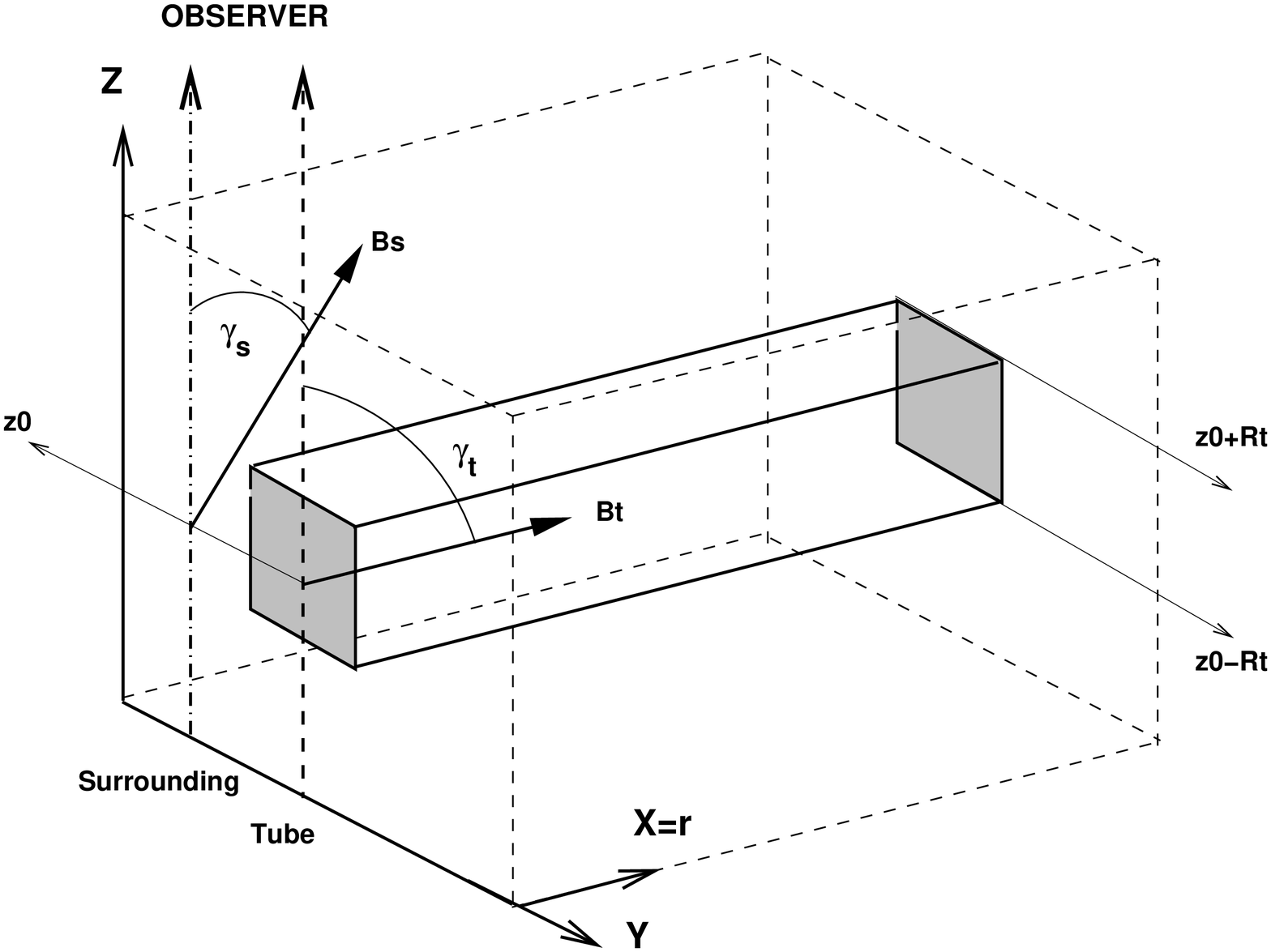}
\end{center}
\caption{Example of the geometrical scheme used in the inversion. 
The radiative transfer equation is solved along the
2 rays (dashed and dot-dashed lines) representing
the embedded flux tube and surrounding atmosphere, respectively.
$\gamma_{\rm s}$ and $\gamma_{\rm t}$ refer to the inclination of
the magnetic field vector with respect to the observer.
In this picture, for simplicity the heliocentric angle
is $\theta=0$ and $\gamma_{\rm t}=90^{\circ}$. The
flux tube structure is sampled using one single ray
(i.e: our model simplifies the shape of the flux tube into
a rectangular cross section).}
\end{figure}

\begin{eqnarray}
\begin{split}
\ve{S}\syn\wl = & \alpha_q \ve{S}_{q}\wl \\ & + (1-\alpha_q) [(1-\alpha_t)\ve{S}_{s}\wl+\alpha_t \ve{S}_{t}\wl]
\end{split}
\end{eqnarray}

\noindent where the indexes 's', 't' and 'q' refer to 
the surrounding, flux tube and quiet Sun (stray light) contributions respectively. The components 
of the Stokes vector are the intensity and polarization profiles \ve{S}$ = (I, Q, U, V)$,
with \ve{S}$_{\rm q} = (I_q, 0, 0, 0)$ being the intensity profile of the observed spectral
lines obtained from the two-component model for the quiet sun from Borrero \& 
Bellot Rubio (2002). $\alpha_{\rm t}$ and $\alpha_{\rm q}$ are the fractional areas
of the resolution element occupied by the flux tube and quiet Sun, respectively.
If the stratification of the physical parameters in the surrounding
atmosphere is denoted as $\ekis_{\rm s}(z)$ (see dot-dashed line in Fig.~2), then
the parameters along the flux tube atmosphere acquire the following form:

\begin{equation}
\ekiss_{\rm t}(z) = \left\{\begin{array}{ll}
  \ekiss_{\rm t} & \rm{if} \;\;\; z \in [z_{0}-R_{\rm t},z_{0}+R_{\rm t}]\\
  \ekiss_{\rm s} & \rm{otherwise}
\end{array} \right.
\end{equation}

\noindent where, as indicated by Fig.~2, $z_0$ is the height where 
the axis of the flux tube is located and $R_{\rm t}$ is its radius.
On the right hand side of Eq.~2, $\ekis_{\rm t}$ is height independent.
Along the surrounding atmosphere $\ekis_{\rm s}$ is also constant with
height, with the exception of the temperature, for all physical parameters. 
For the temperature stratification T$_{\rm s}(z)$ we have chosen, 
as in Paper II, the penumbral model of Del Toro Iniesta et al. (1994).
Note that Eq.~2 ensures that the physical stratifications along
the flux tube atmosphere (vertical dashed line in Fig.~2) are the
same as in the surrounding atmosphere above and beneath the
flux tube: $z<z_o-R_{\rm t}$ and $z>z_o+R_{\rm t}$. At the lower
and upper boundaries of the flux tube ($z^{*}=z_0 \pm R_{\rm t}$) 
the physical quantities encounter a sharp transition
of magnitude $\Delta \ekis=\ekis_{\rm t}(z^{*})-\ekis_{\rm s}(z^{*})$.
These jumps/gradients are the essential ingredients of this model to explain the 
net circular polarization (NCP) observed in the sunspot penumbra 
(S\'anchez Almeida \& Lites 1992; Solanki \& Montavon 1993; cf. Mart{\'\i}nez 
Pillet 2000). Finally, we shall mention that both the surrounding magnetic
atmosphere and the flux tube are forced to be in total pressure balance
at each height (see Sect.~2 in Paper II for details).

Once the synthetic profiles are computed for an initial guess of free parameters,
a merit function is constructed by comparing with the observed penumbral profiles:

\begin{equation}
\chi^2 = \frac{1}{L} \sum_{j=1}^{4} \sum_{k=1}^{M} [S_{\rm obs}^{j}(\lambda_k)
-S_{\rm syn}^{i}(\lambda_k)]^2 w_{j}^2
\end{equation}

In this equation, $L$ stands for the total number of free parameters
in the uncombed model (i.e.:17, see paper II). The index $j=1,...,4$ denotes 
the four components of the Stokes vector, while $k$ samples 
each profile in the wavelength direction. Finally, the weighting factor $w_{j}$
is used to favor some spectral lines or polarization states.
The weights we have used in our inversions are indicated in Table~2. We 
are giving special attention to the Stokes $V$ signal for the 
neutral iron lines at 6301.5 and 6302.5 \AA. This is because we are specially 
interested in extracting the information contained in the net circular polarization
(i.e.: gradients along the line of sight in the physical parameters, 
radius of the penumbral flux tubes, etc.).
In addition, some extra weight has been given to the total intensity signal, Stokes $I$,
of the Ti I line at 6303.7 \AA. The combination of the very small 
excitation energy of this line and the relatively low ionization energy of
Titanium, makes it fairly sensitive to variations in the 
temperature stratification. In fact, the Ti I line behaves similarly to
the OH lines blended with the Fe I 15652.5 \AA~ line (see Paper II), 
displaying a high equivalent width in the umbra and at the umbral-penumbral boundary, which
rapidly decreases at larger radial distances from the center of the
sunspot. The polarization profiles of the last two lines in Table~2 are
small compared to those of the first two Fe I lines. Therefore, they are 
more affected by noise and are of relatively minor importance during the inversion.

\begin{table}
\caption[]{Relative weights, $w_{j}$, entering in the merit function $\chi^2$,
for each spectral line and each component of the Stokes vector.}
\begin{center}
\tabcolsep .6em
\begin{tabular}{lccccc}
\hline
Species & $\lambda_0$ & $w_i$ & $w_q$ & $w_u$ & $w_v$\\\hline
\hline
Fe I & 6301.5012 &  1  & 1  &  1  &  8\\           
Fe I & 6302.4916 &  1  & 1  &  1  &  8\\
Fe I & 6303.4600 &  1  & 0.2  &  0.2  &  0.2\\           
Ti I & 6303.7560 &  4  & 0.2  &  0.2  &  0.2\\
\hline
\end{tabular}
\end{center}
\end{table}

Besides the $\chi^2$ function from Eq.~3, its derivatives
with respect to the free parameters of the problem are computed
numerically. A Levenberg-Marquadrt non-linear inversion algorithm
 is fed with these ingredients in order to iteratively
modify the initial guess set of parameters, \ekis$_{0}$~ (guess
magnetic field vector, guess velocity), until a minimum
in the merit function is reached. The inversion code described
by Frutiger et al. (1999, 2000) and Frutiger (2000) is employed.

\section{Inversion results}%

We have selected for the inversion the region enclosed by the square in Fig.~1.
It lies on the limb side of the penumbra, covers an area of about 15$\times$15
arc sec and encloses the line of symmetry of the sunspot. This region was selected
because, as can be seen in Fig.~1 (bottom panel), there is a maximum in the
area asymmetry of the circular polarization profiles, and hence it seems very
suitable to constrain the values for the flux tube
radius. The profiles in each spatial pixel are inverted individually starting
always from the same set of guess parameters. 
Let us first give some attention to the quality of the fits
produced by the uncombed model. In Fig.~3 we present some examples of 
the observed and fitted Stokes $V$
profiles for three pixels located at increasing distance from the sunspot
center. The fits to the observed data are reasonably accurate, 
with differences normally well below 0.5\%~ of the continuum intensity. Even Stokes 
$V$ profiles with typical shapes of Stokes $Q$ and $U$ (see S\'anchez Almeida
\& Lites 1992), are almost perfectly reproduced (Fig.~3; middle panels),
as are single-lobed Stokes $V$ profiles (Fig.~3; bottom panels).
Note that, in contrast to the Fe I lines at 1.56 $\mu$m,
these fits cannot be achieved by simple 2 component models that neglect gradients
along the line of sight in the magnetic field vector and velocity
(see Papers I and II). The reason is that the produced area asymmetry with
such models would be zero,
giving rise to Stokes $V$ profiles that
will have equal positive and negative areas (although they may have several
lobes due to the cross-over effect) in strong
contrast to the observed profiles (see Fig.~3).

\begin{figure*}
\begin{center}
\includegraphics[width=9cm]{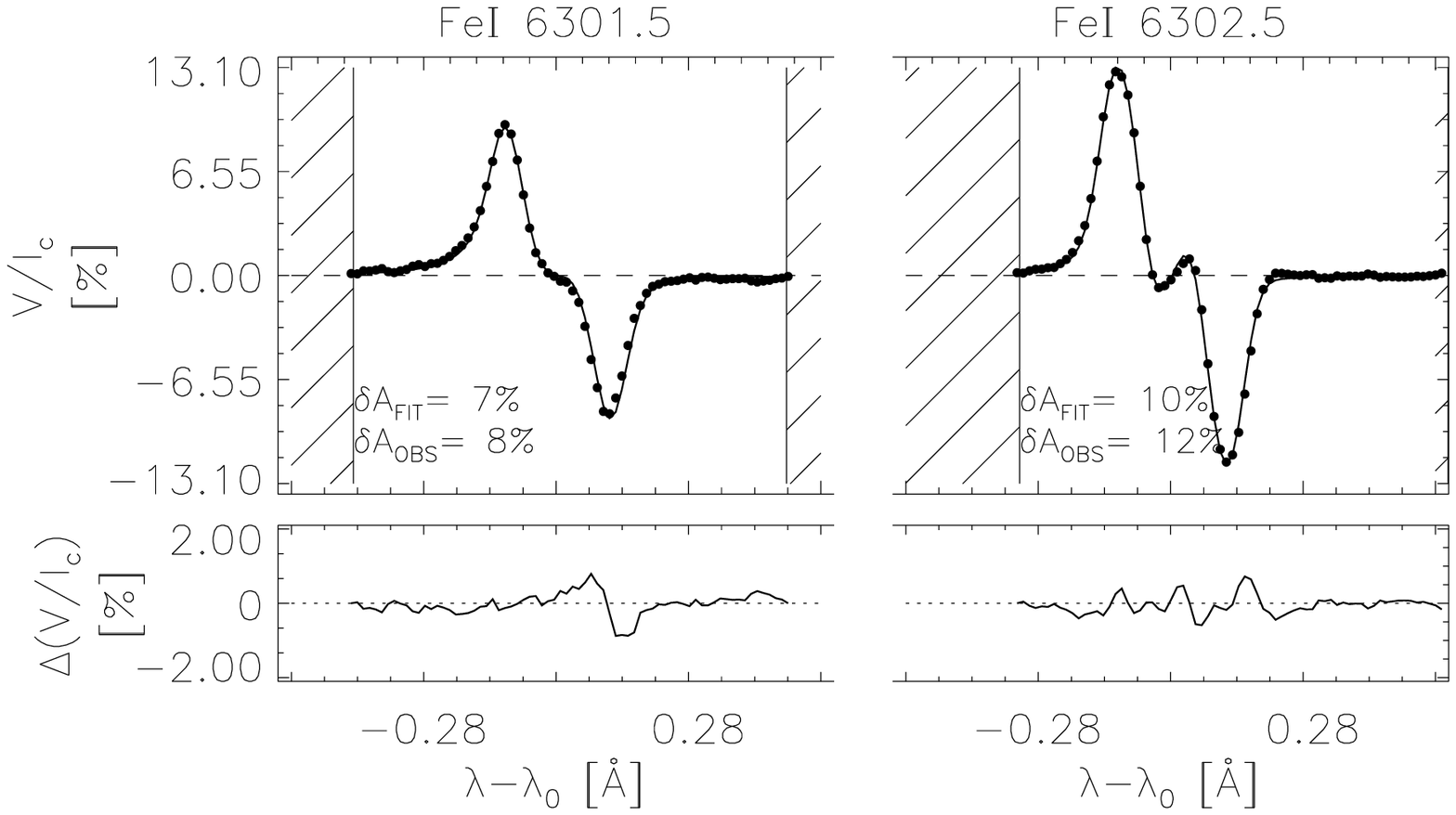}
\includegraphics[width=9cm]{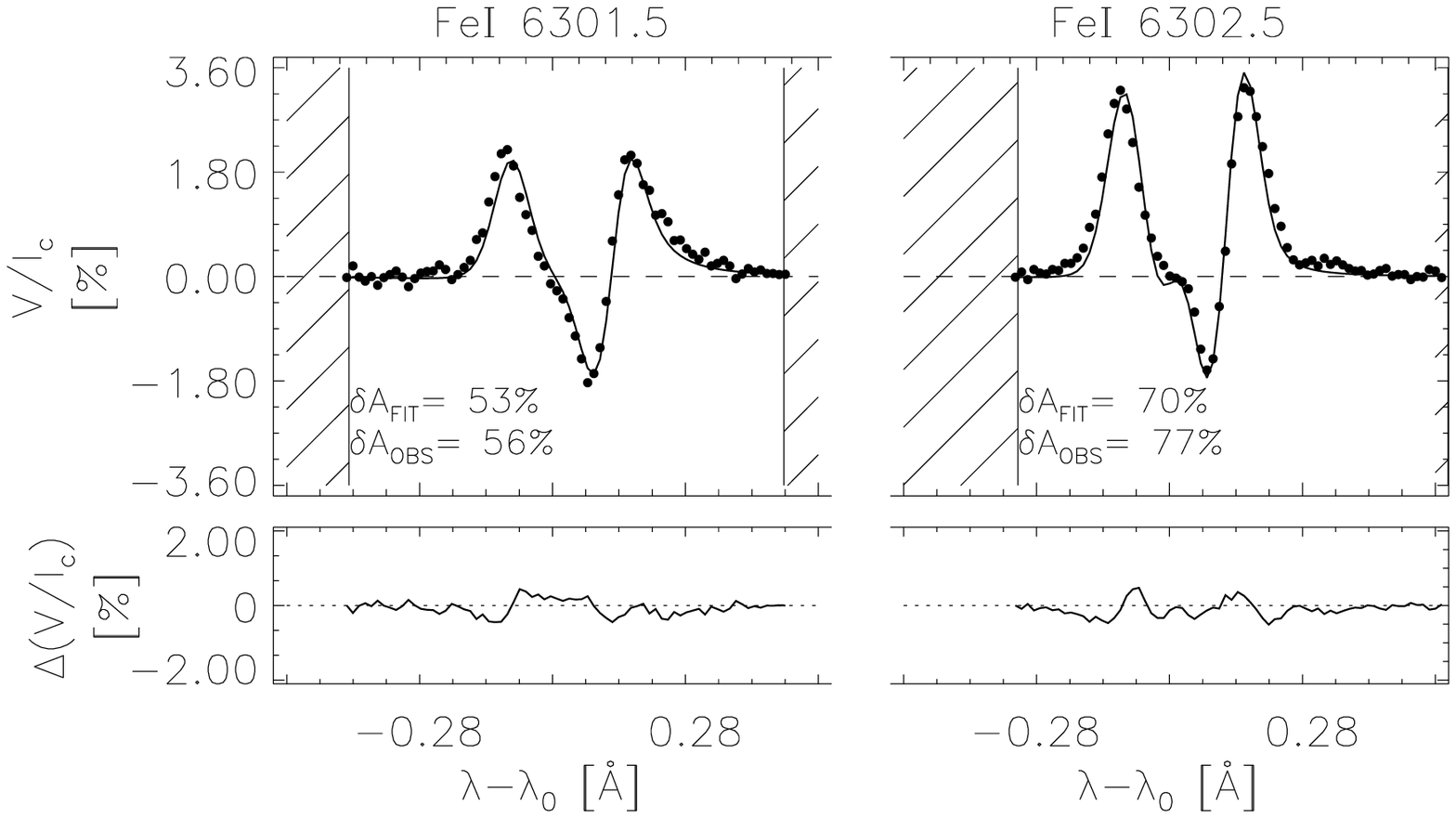}
\includegraphics[width=9cm]{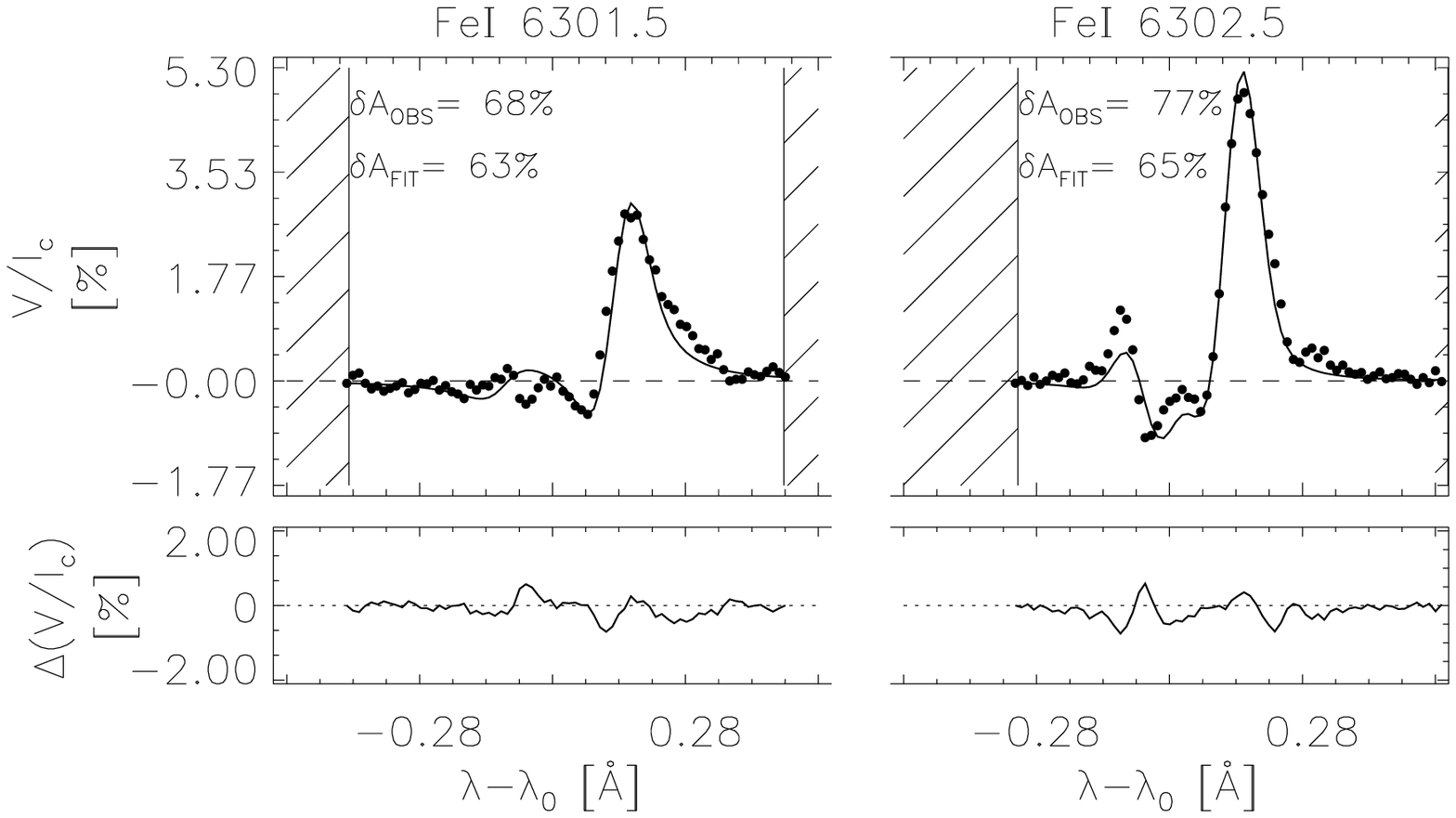}
\end{center}
\caption{Three examples of observed (filled circles) and fitted (solid lines) 
  Stokes $V$ profiles for Fe I 6301.5 (left panels) and Fe I 6302.5 \AA~
  (right panels) lines. These profiles have been taken from three locations 
  with increasing radial distance: $r/R=0.45,0.65,0.85$, such that the upper 
  panel corresponds to the inner penumbra and the bottom panel lies at the
  outer boundary of the sunspot. The observed and fitted Stokes $V$ area
  asymmetry, $\delta A$ are also indicated.}
\end{figure*}

As a result of the inversion, the stratifications of the different physical parameters
with height in the atmosphere were obtained for both the flux tube
component (dashed ray path in Fig.~2) and the surrounding atmosphere (dot-dashed
ray path in Fig.~2). We have extracted, for all inverted pixels, the values of the 
physical parameters (temperature, LOS velocity, magnetic field strength 
and inclination, gas pressure etc.) at $z=z_0$ and plotted them in Fig.~4, as a function 
of the normalized radial distance in the penumbra $r/R$, with $R$ being the radius
of the sunspot as defined by the outermost contour in Fig~.1. For all those parameters, 
except for the temperature, we plot separately the radial behavior 
along the flux tube (dashed lines and light gray shaded area) and its surrounding 
atmosphere (dot-dashed line and dark gray shaded area). The dashed and dot-dashed
lines indicate the average properties, at a given radial distance,
of the flux tubes and surrounding atmosphere, while the shaded areas
indicate the maximum deviations from the average values obtained
from all inverted points at the same radial distance.
For the temperature we plot (solid line) the difference between 
the flux tubes and the magnetic surrounding atmosphere. The 
filling factor of the flux tube component, $\alpha_t$ (dashed line in Fig.~4; 
bottom left panel) corresponds, obviously to the flux tube alone, 
with $1-\alpha_t$ being the area covered by the
surrounding atmosphere (Eq.~1). In this same plot the amount of stray light
contamination $\alpha_{\rm q}$ is represented by the solid line. We retrieve
consistent values with those found by Lites et al. (1993) who analyzed
ASP data of another sunspot.

As in previous investigations (Borrero et al. 2004, 2005), 
we have transformed the magnetic field inclination (middle left panel in Fig.~4), 
that is always obtained in the observer's reference frame from the inversion, 
into the local reference frame. Therefore, in the following we discuss the 
zenith angles $\zeta$ instead of the observer frame inclination $\gamma$, 
with $\zeta=90^{\circ}$ meaning that the 
magnetic field is parallel to the solar surface (marked with a horizontal 
dotted line in the middle left panel of Fig.~4). 
The lower right panel in Fig.~4 shows the gas pressure along the
penumbral flux tubes (dashed line) and surrounding atmosphere (dot-dashed line)
at $z=z_0$. The gas pressure difference between these two components is indicated
by the solid line. As imposed, $\Delta P_{\rm gas} \propto \Delta B^2$ since 
at the same height both atmospheres are in lateral pressure balance 
(see Eq.~4 in Paper II).

\begin{figure*}
\begin{center}
\begin{tabular}{cc}
\includegraphics[width=7cm]{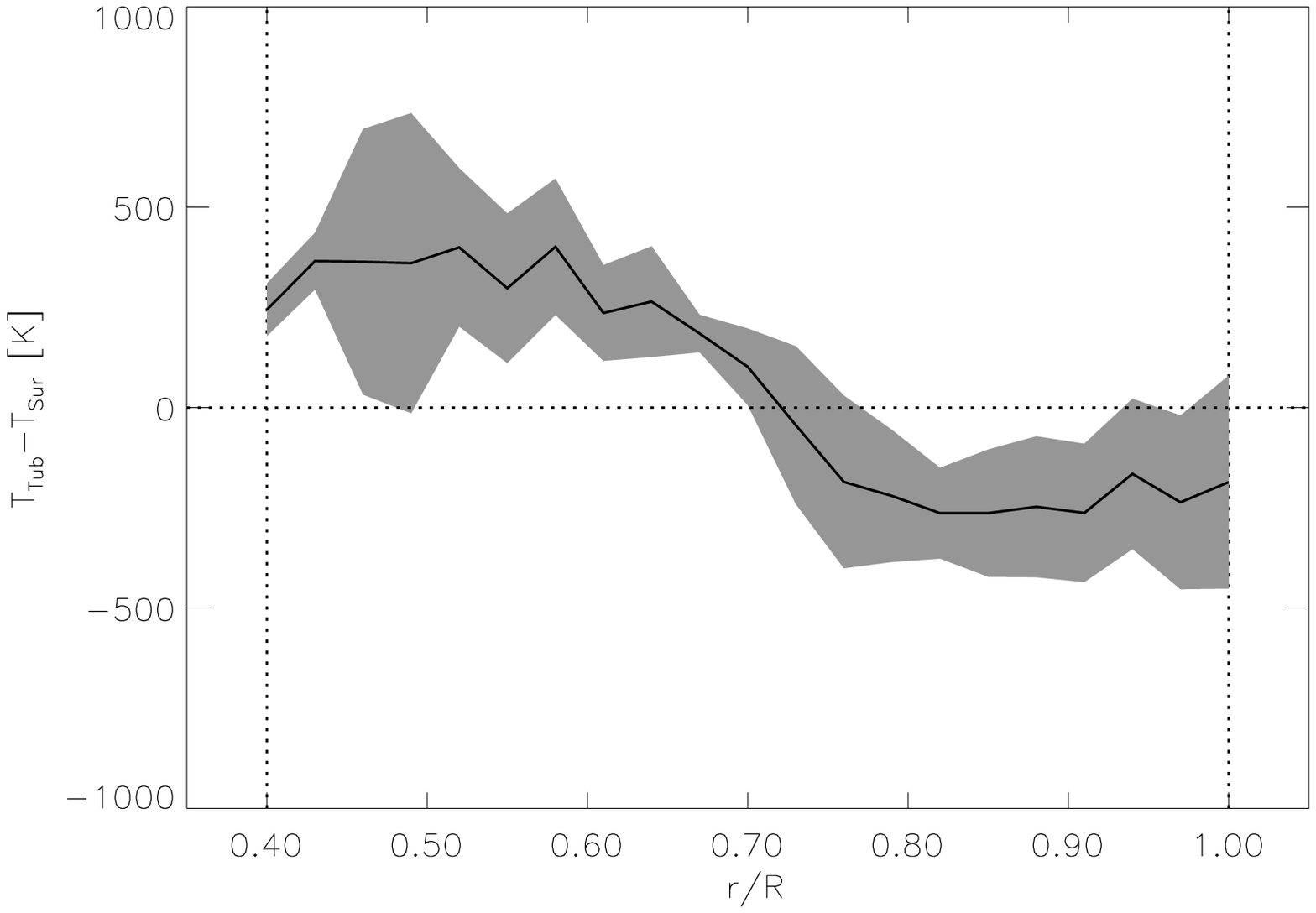} &
\includegraphics[width=7cm]{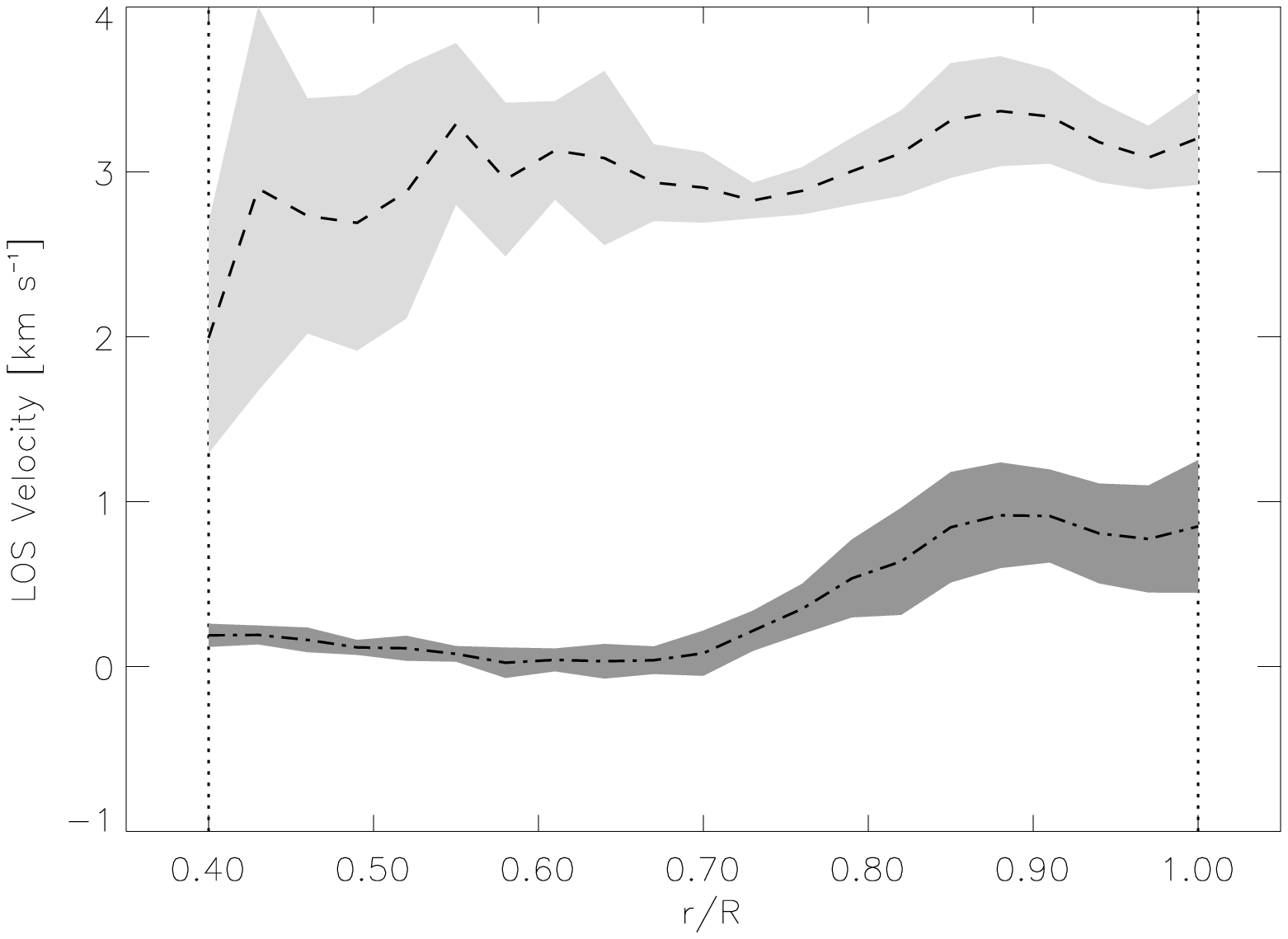} \\
\includegraphics[width=7cm]{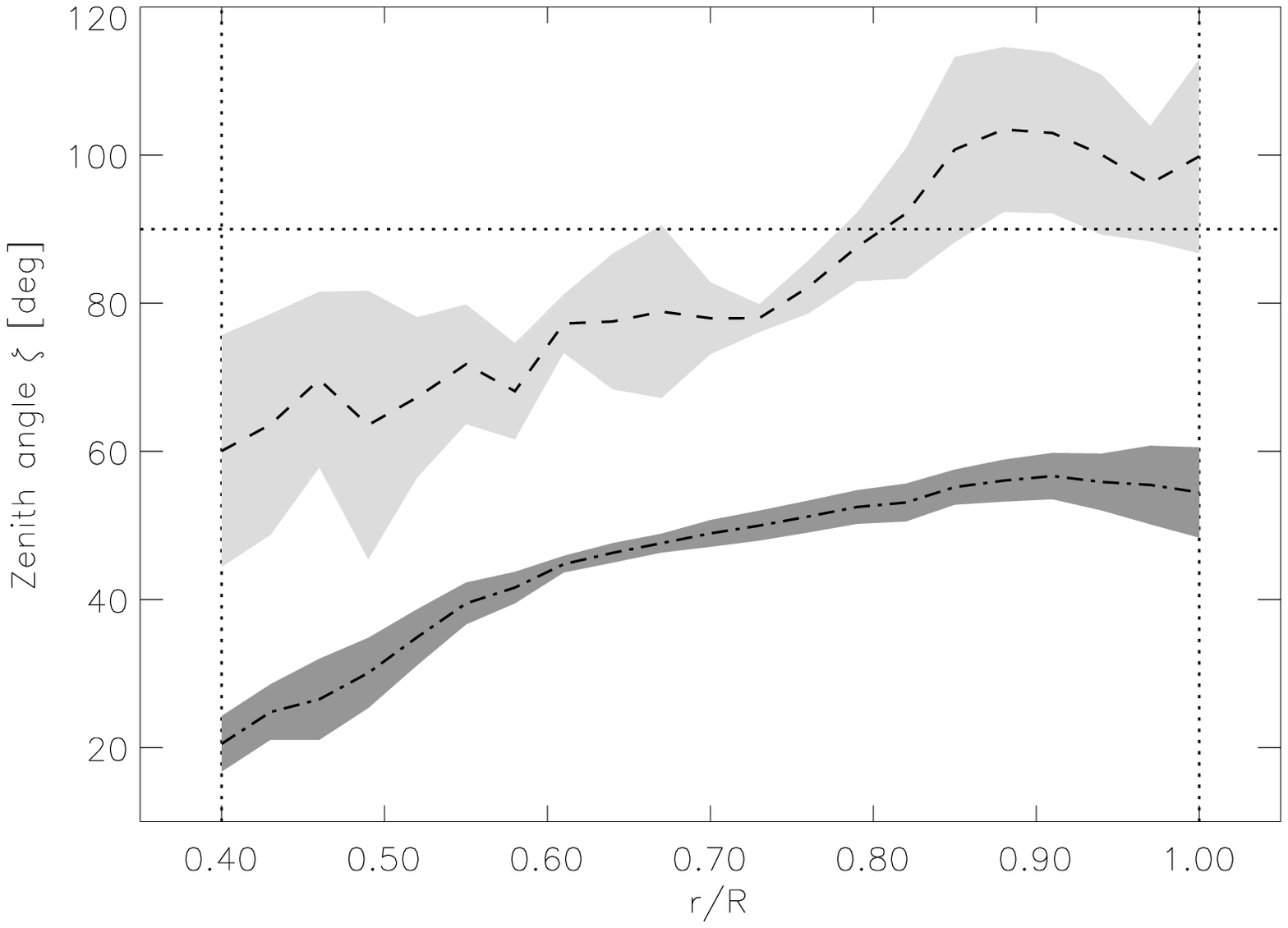} &
\includegraphics[width=7cm]{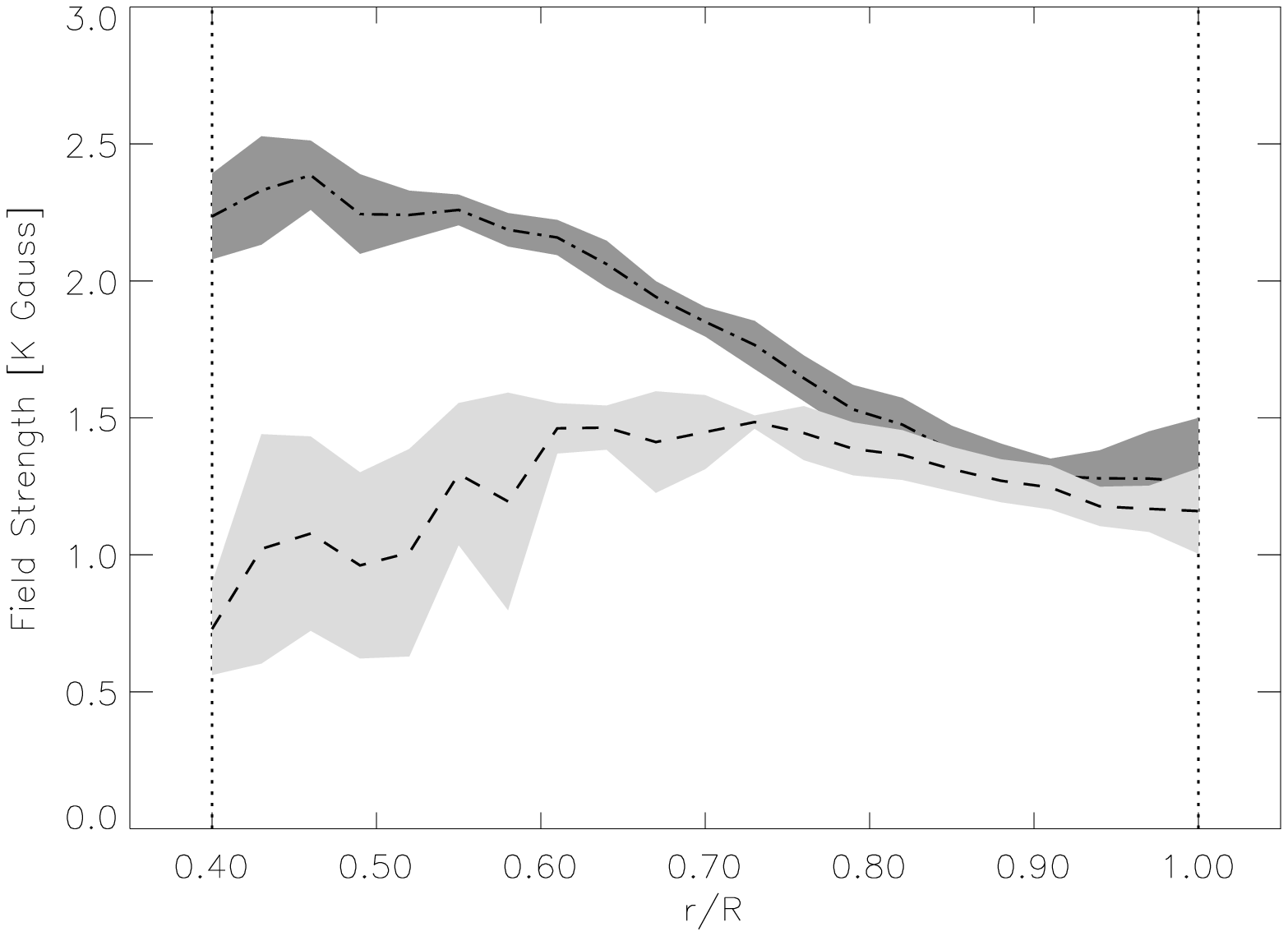} \\
\includegraphics[width=7cm]{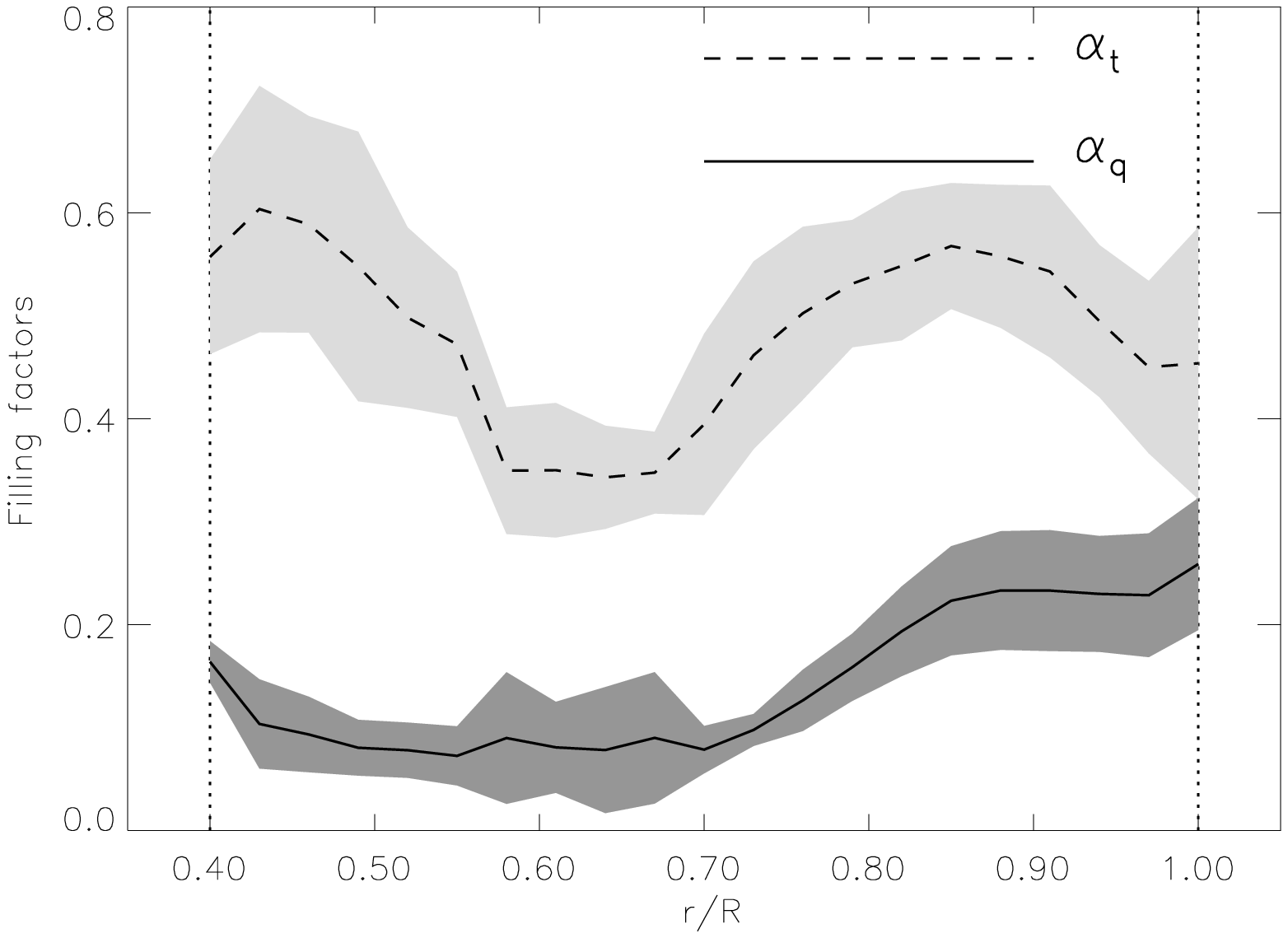} &
\includegraphics[width=7cm]{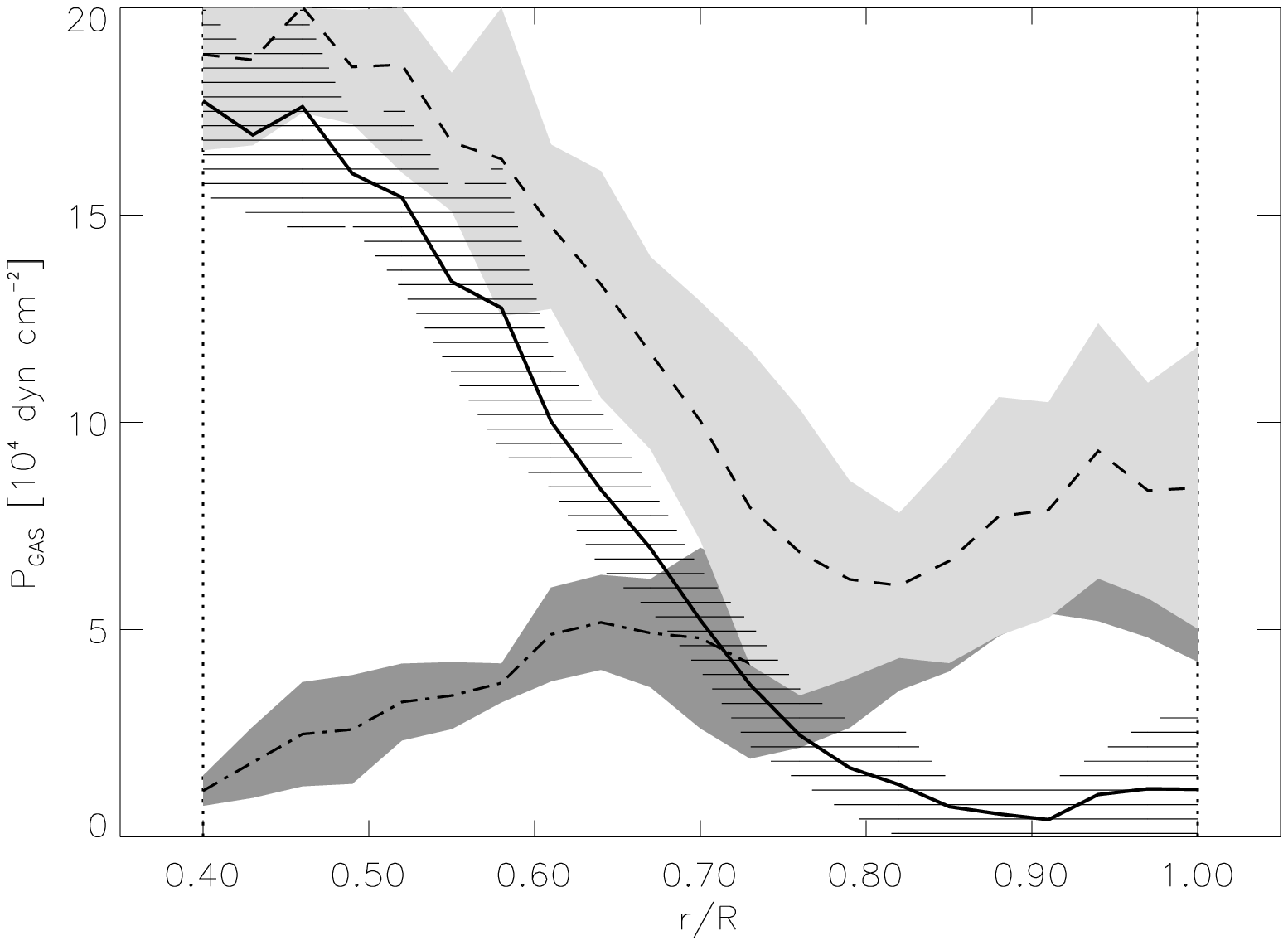}
\end{tabular}
\end{center}
\caption{Radial variation of the properties of the magnetic surrounding 
(dot-dashed lines) and the penumbral flux tube (dashed lines). Shaded areas
represent the maximum deviations around the mean obtained from the inversion
of all pixels at a given radial distance. From left to right and top to bottom:
temperature difference between the flux tube and its surroundings, LOS
velocity, magnetic zenith angle $\zeta$, magnetic field strength $B$, flux tube filling
factor $\alpha_t$ (dashed line) and fraction of stray light $\alpha_{\rm q}$ (solid line),
and gas pressure along the flux tubes (dashed), surrounding atmosphere (dot-dashed)
and difference between them (solid line and area with horizontal stripes).
All quantities have been taken at a geometrical depth that
corresponds to the central position of the flux tube: $z=z_0$.}
\end{figure*}

\subsection{Discussion}%

A comparison of the results plotted in Fig.~4 with those
obtained with the Fe I 1.56 $\mu$m lines by means
of the uncombed model (see Paper II; Figs.~5 and 6) shows very similar
radial trends. It is gratifying to see that the same model gives
so similar results when applied to a set of spectral lines with very different 
atomic properties and thermodynamic sensitivities. In particular:

\begin{itemize}
\item Flux tubes appear hotter than the surrounding atmosphere 
in the inner penumbra, supporting our previous results from the inversion
of the Fe I lines at 1.56 $\mu$m (see Fig.~3; top left panel).
\item Flux tubes are inclined with respect to the vector normal to the solar
surface in the inner penumbra, $\zeta_{\rm tub} \sim 60^{\circ}$ and
become horizontal $\zeta_{\rm tub} \sim 90^{\circ}$ in the middle penumbra
(see Fig.~3; middle left panel). At the outer penumbra $r/R > 0.8$
for most spatial pixels the flux tubes start to bend down, $\zeta_{\rm tub} > 90^{\circ}$, as if
returning into the solar interior. Very similar radial dependences were found in Papers I and II. 
\item The line of sight velocity
along the flux tubes (dashed line in Fig.~3; top right panel) is already 
significant at the inner penumbra $v_{\rm los} \sim
2$ km s$^{-1}$. It increases slightly up to $v_{\rm los} \sim 3$ km s$^{-1}$ and
remains fairly constant for $r/R > 0.5$. We do not detect here any abrupt
decrease of the flow speed at large radial distances as seen in Paper II.
There is also no sign of any enhancement in the flux tube temperature
in the outer penumbra. This seems to imply that the signatures of shock fronts due to
supercritical speeds reported by Borrero et al. (2005) are not universally seen
in sunspot penumbrae. Indeed, the sunspot analyzed in this work is 
closer to the center of the solar disk than that analyzed in Paper II,
and therefore, the LOS velocities obtained here are less representative of the true
velocities. Results at smaller radial distances are however, completely consistent
with those obtained from the inversion of the Fe I lines at 1.56 $\mu$m.
In particular, the high Evershed velocities found in the inner penumbra
(Fig.~4, top right panel), suggest that the penumbral flux tubes
emerge into the photosphere from subphotospheric layers already
carrying high velocities. This agrees with the model results of Schlichemaier
et al. (1998a, 1998a, 1999) and Schmidt \& Schlichenmaier (2000).
\item The magnetic field inclination and line of sight velocities found in the
surrounding component are also compatible with (in fact, they are strikingly
similar to) those obtained in Papers I and II (see e.g. Fig.~5 and 6 in
Borrero et al. 2005).
\item The field strength in the surrounding atmosphere rapidly decays 
radially, while along the flux tube it seems to
increase at first until $r/R = 0.6$ and slowly decreases afterward.
Note however that if these averaged flux tube properties
would correspond to a single flux tube that crosses the penumbra,
the magnetic field in the inner footpoint, $B \sim 750$ G would be smaller than in the
outer footpoint, $B \sim 1100$ G. Such a radially
increasing magnetic field is in good agreement with the requirement of
the siphon flow mechanism to drive the Evershed flow in an homogeneous penumbra
(see Montesinos \& Thomas 1997 and references therein). This particular behavior
must be considered with some caution however (the Fe I lines at 1.56 $\mu$m did not reveal
this behavior in Papers I and II; see also next paragraph).
The main result is still consistent with our
previous work, that is: while the surrounding magnetic field decreases radially
very rapidly, the field strength along the flux tube exhibits a much smaller
and slower variation, being $B_{\rm sur} >> B_{\rm tub}$ in the inner penumbra, but
$B_{\rm sur} \sim B_{\rm tub}$ at the outer sunspot boundary. Since
the radial dependence of the gas pressure difference at $z=z_0$
between the flux tubes and their magnetic surroundings scales as
$\Delta B^2$ this implies, in agreement with Papers I and II,
that the gas pressure difference decreases abruptly with radial
distance in the penumbra (see Fig.~3; bottom right panel). This 
gas pressure excess in the flux tubes at the inner penumbra is of course
consistent with the higher temperatures found there.
\item The individual behaviour of the gas pressure in the penumbral 
flux tubes and their magnetic surroundings (Fig.~4; bottom right panel)
is also instructive. Whereas the gas pressure in the flux tube's surroundings
slowly increases outwards, along the flux tube it displays a rapid drop.
In principle, such a pressure gradient as seen in the flux tubes should
suffy to drive a siphon flow. However, as
already mentioned in Paper II, such a conclussion is correct only if the flux tube
is located at the same absolute geometrical height in the inner and the outer penumbra. 
Since the central position of the flux tube $z_0$ may change radially
if the Wilson depression were to depend on $r/R$, we cannot conclude that what we see
is actually a siphon flow (although the results are completely consistent with one). 
However, we can the argument around and assume that this is the case. Then we can
estimate by how much thee flux tubes have risen from the inner to the outer
penumbra: $z_0(r/R=1.0)-z_0(r/R=0.4) \lesssim 100$ Km. These results are only slightly
smaller than those found by Solanki et al. (1993) and Mathew et al. (2004)
\item The properties obtained for the flux tube atmosphere are always
less accurate that those for the magnetic atmosphere that surrounds
them, in particular in the inner penumbra, where the light gray 
shaded areas indicating the deviations with respect to the average
flux tubes properties are largest. This indicates caution 
when interpreting our results for $r/R < 0.5$. The reason for
this is to be found in the information contained in the polarization
signals near the umbra, where for example, the Stokes $V$ profiles do not
show any particular peculiarities (see Fig.~3; top panels) or
distinctive signal of the presence of two different components within
the resolution element.
\end{itemize}

\subsection{Vertical extension of the Penumbral Flux tubes}%

Already in Papers I and II, where the umcombed model was applied to
the Fe I lines at 15648 and 15652 \AA~, we pointed out that their
visible counterparts (Fe I lines at 6300 \AA), being far more sensitive to 
the vertical stratification in the physical parameters, might help to constraint
 the values for the vertical extension of the penumbral fibrils. In this section 
we will investigate this issue.

In order to produce a non vanishing net circular polarization an atmospheric
model must include gradients along the line-of-sight in the velocity and
magnetic field vector (Landolfi \& Landi Degl'Innocenti 1996). In the case of
the penumbra of sunspots, the net circular polarization observed in the Fe I lines
at 6300 \AA~ is so large that gradients in the inclination of the magnetic
field and line-of-sight velocity as large as $\Delta \gamma \simeq 45
^{\circ}$ and $\Delta \rm v_{los} \simeq 1.5$ km s$^{-1}$ must be introduced
(S\'anchez Almeida \& Lites 1992). 

Such gradients would produce large curvature forces and electric currents 
that are difficult to match with the idea of a sunspot in hydrostatic equilibrium 
(Solanki et al. 1993). A way out of this problem was proposed by Solanki \& Montavon 
(1993) who realized that including a three layered atmospheric model where the
middle one would have larger velocities and a more horizontal magnetic field
could reproduce the observed NCP with no need to invoke net (i.e. large scale) gradients
along the line of sight, but rather strong gradients on a small scale that are
compensated as the line-of-sight crosses the three atmospheric layers. This is
the basic idea of the {\it uncombed model}. In our representation of the uncombed
model these three layers are identified with the different
stratifications along the dashed ray path in Fig.~2. We have already 
identified the intermediate layer as a horizontal flux tube (carrying the
Evershed flow) which is embedded in a magnetic surrounding where the magnetic
field is more vertical and essentially at rest. In addition, we are considering
a pure surrounding atmosphere (dot-dashed line in Fig.~2; see also 
Mart{\'\i}nez Pillet 2000). Note that since the physical
parameters along the surrounding atmosphere are constant with height (see
Sect.~3) no net circular polarization is produced here. The NCP is exclusively
produced by the ray that crosses the flux tube due to the strong gradients
present at the tube's upper and lower boundaries: $z^{*}=z_0 \pm R_{\rm t}$.

This realization of the uncombed model has been applied in Paper II 
to penumbral spectropolarimetric data observed in the Fe I lines 
at 1.56 $\mu$m. This allowed us to characterize the
horizontal inhomogeneities of the penumbral fine structure, but we were not
able to model the vertical ones. Since these lines show only small 
amounts of NCP (with typical values for the area
asymmetry of $\delta A \sim 3-5$ \%), the positions of the
boundary layers where the gradients along the line of sight are produced, 
therefore the thickness of the penumbral flux tubes, were not properly constrained
in the inversion process\footnote{In this paper we will refer both to
the NCP and Area Asymmetry as if they were the same. In fact, the second is 
related to the first one by normalizing to the total (unsigned) area of Stokes $V$.
This is of no consequence for our present discussion}.

\begin{figure}
\begin{center}
\includegraphics[width=8.5cm]{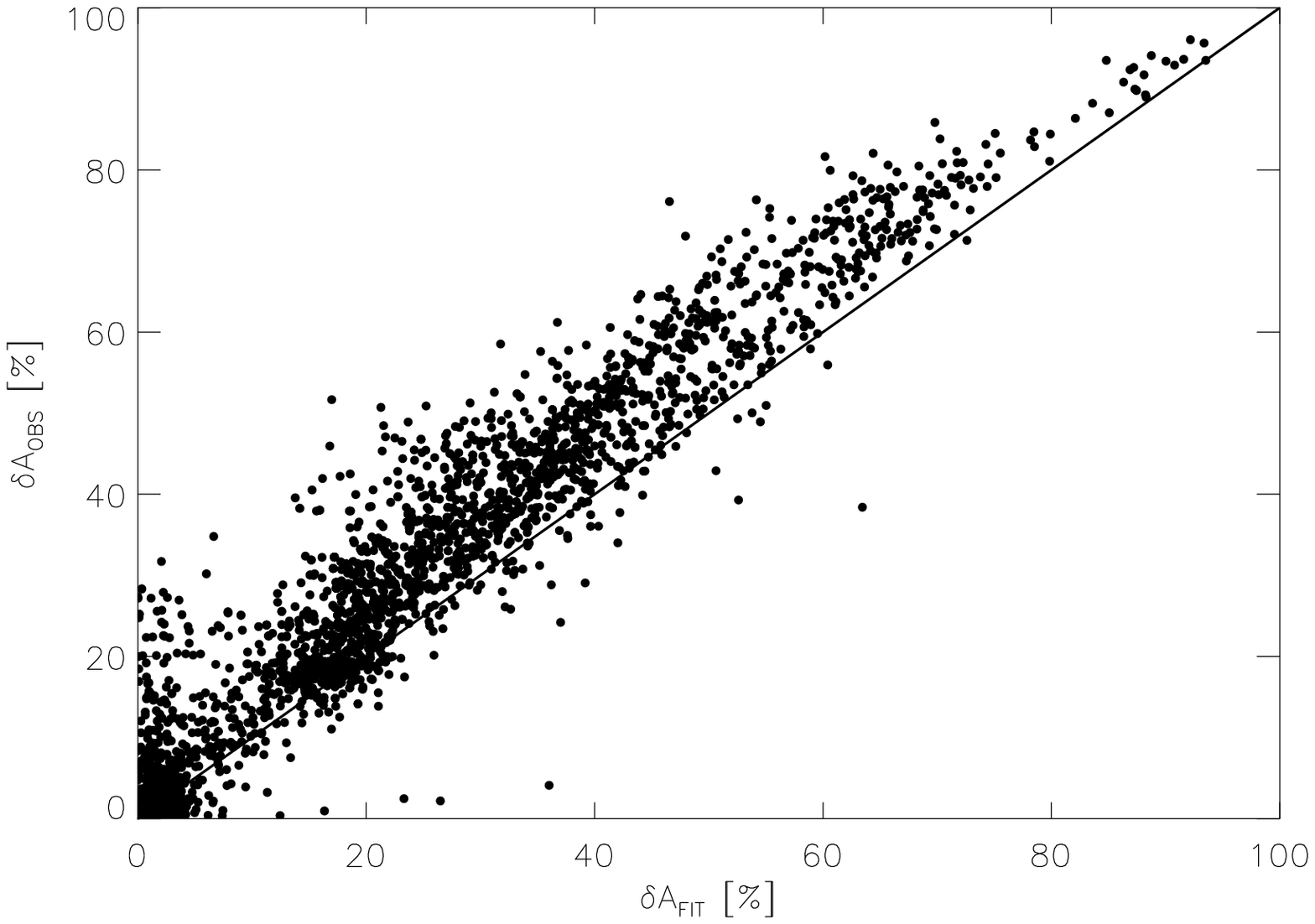}
\includegraphics[width=8.5cm]{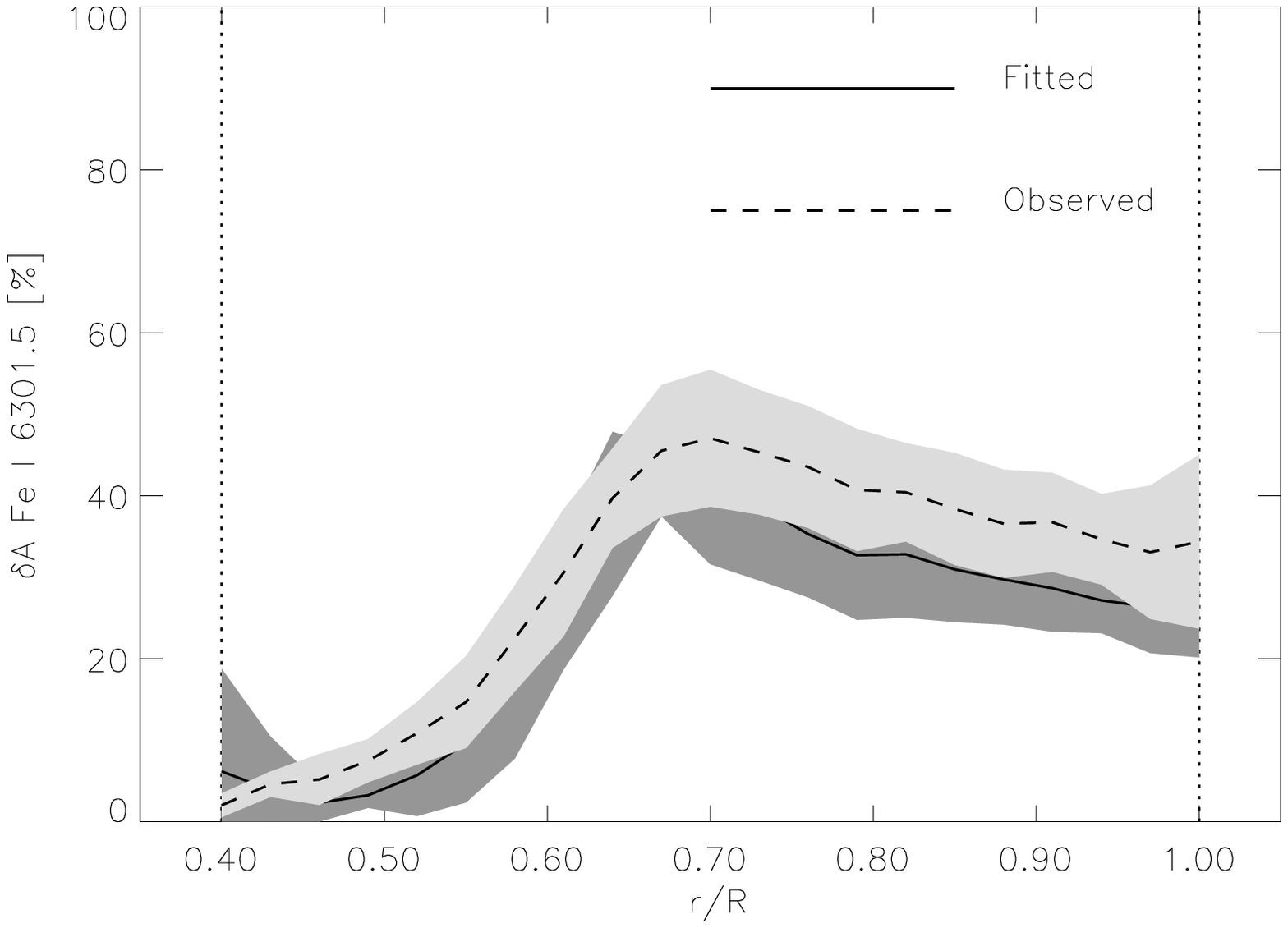}
\includegraphics[width=8.5cm]{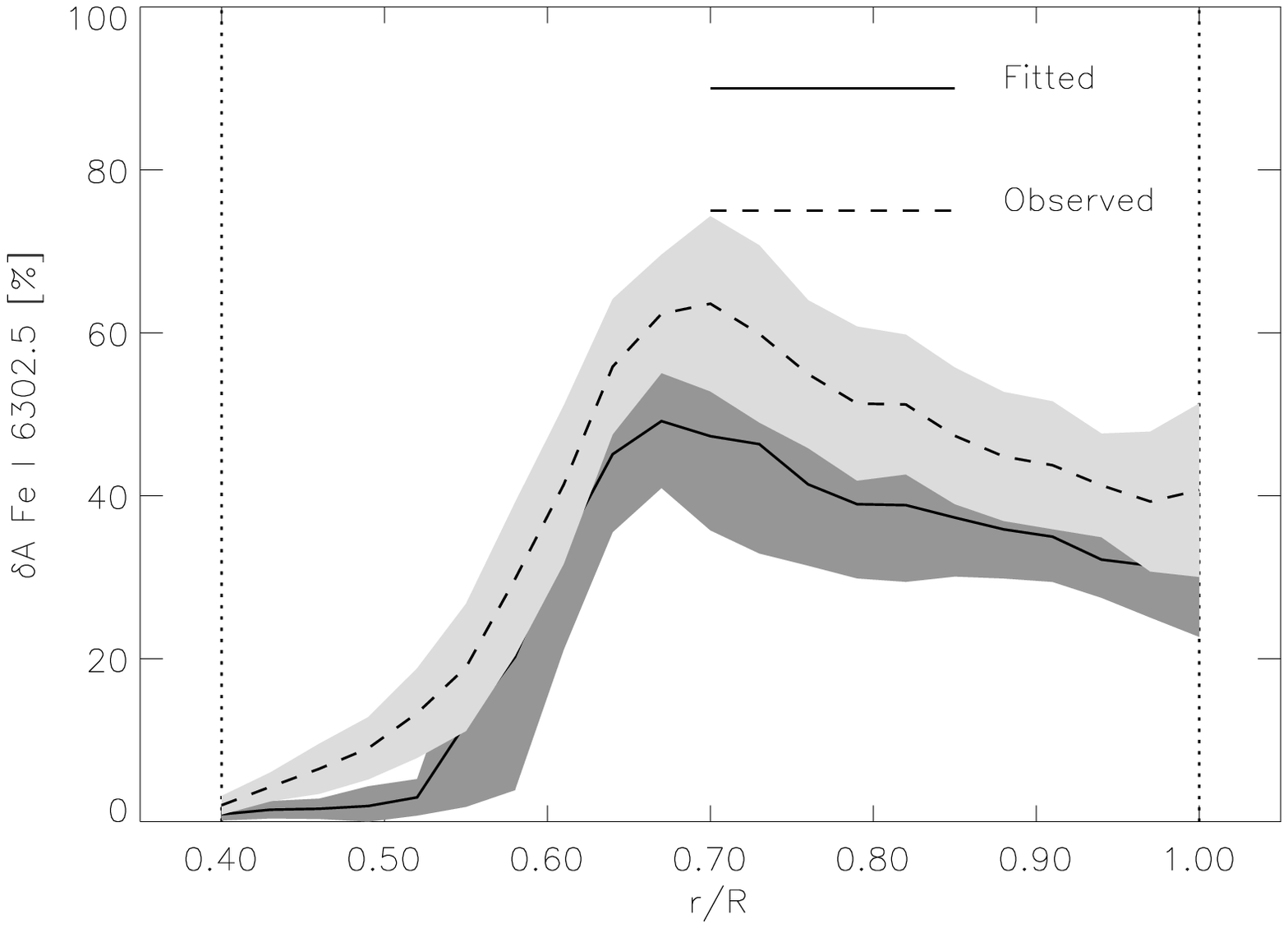} 
\end{center}
\caption{Top panel: Scatter plot of fitted Stokes $V$ area asymmetry versus observed
  values. The straight line indicates the expectation values. Middle panel: radial
  variation of the area symmetry of the observed (dashed line) and fitted (solid line)
  Fe I 6301.5 \AA~ Stokes $V$ profiles. Shaded areas corresponds to the maximum and 
  minimum deviations from the mean. Bottom panel: same,
  but for Fe I 6302.5 \AA.}
\end{figure}

\begin{figure}
\begin{center}
\includegraphics[width=8cm]{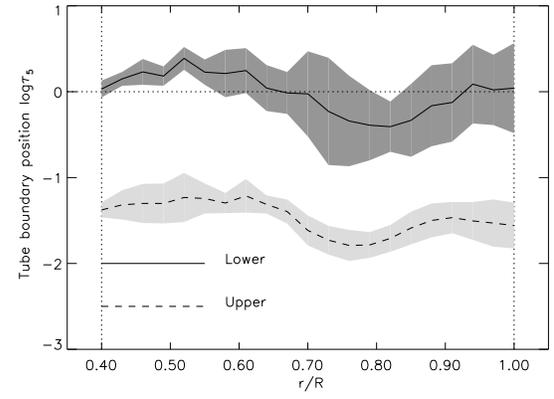}
\includegraphics[width=8cm]{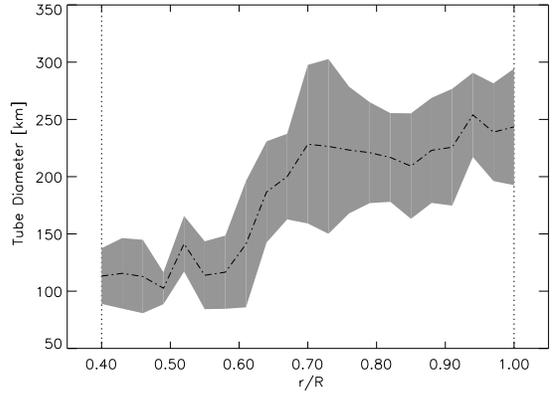} 
\end{center}
\caption{Top panel: locations, in the optical depth scale, of the lower (solid
  line) and upper (dashed line) boundaries of the flux tube as a function of
  the radial distance in the penumbra. Bottom panel: inferred flux tube
  diameter as a function of $r/R$.}
\end{figure}

\begin{figure}
\begin{center}
\includegraphics[width=8.5cm]{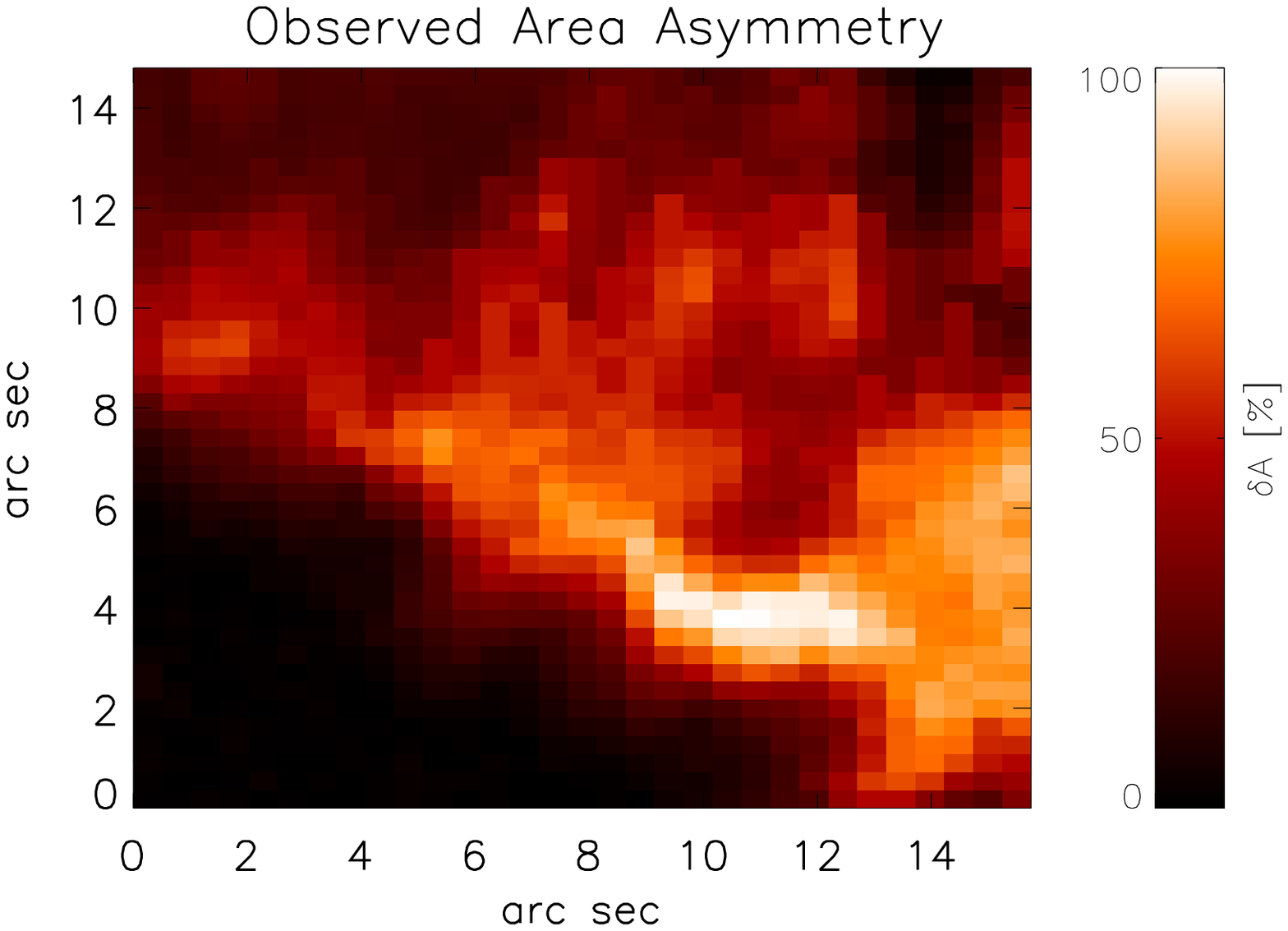}
\includegraphics[width=8.5cm]{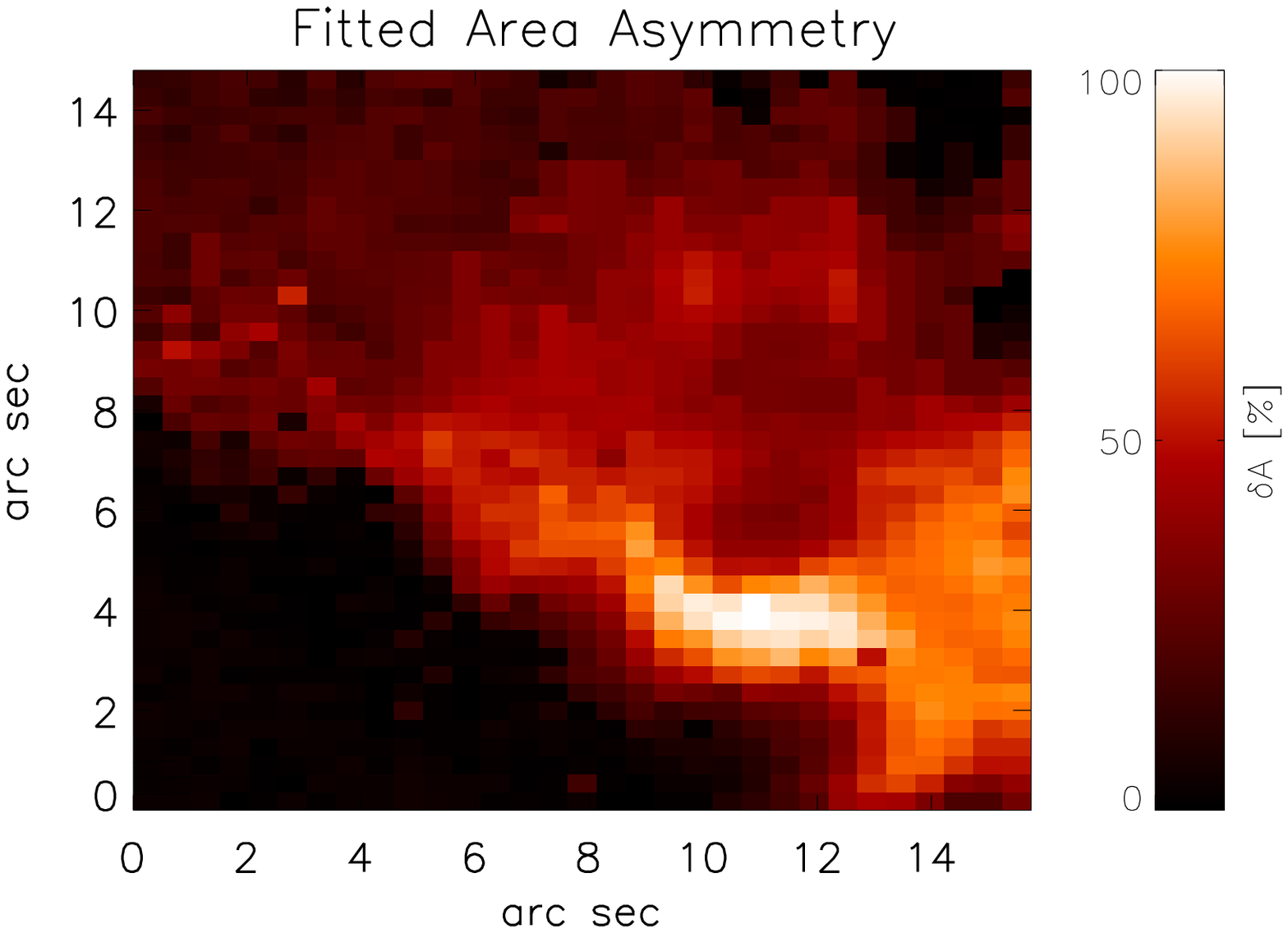}
\end{center}
\caption{Top panel: Map of the area asymmetry  
in the white square box in Fig.~1. {\it Top panel}: observed area
asymmetry. {\it Bottom panel}: area asymmetry produced by the uncombed model.}
\end{figure}

The area asymmetry in the Fe I lines at 6300 \AA~ is so large that the
vertical structure (i.e. discontinuities along the line-of-sight produced at the flux
tube boundaries) of the magnetic field and velocity vectors play
the leading role. Fig.~5 (top panels) demonstrates that the uncombed model
now reproduces the large observed area asymmetry of the visible 
Fe I lines very well (cf. Fig.~ 7 in Paper I or top panel of Fig.~7 in Paper II; see
also Fig.~18 in S\'anchez Almeida 2005). As a function of the radial distance 
(Fig.~5; middle and bottom panels) we see that the observed area 
asymmetry (dashed lines) increases radially up to $\delta A \sim 50-70$ \% in the
middle penumbra ($r/R \sim 0.7$) and decreases slowly hereafter. The area
asymmetry of the fitted profiles (solid lines) mimics this behavior, although
it slightly underestimates the observed values. 

Note that the radial behavior of both the observed and the fitted
NCP follow the prediction of Solanki \& Montavon (1993) in their Fig.~4.
They realized than even though the amount of NCP generated at the flux tube's
boundaries depends on the magnitude of the jump in the velocity and magnetic
field inclination at that location, the NCP displayed a maximum where the
average line of sight magnetic field between the flux tubes and their magnetic
surroundings, $\tilde{\gamma} = \frac{1}{2}[\gamma_{\rm tub}+\gamma_{\rm sur}]$,
was close to 90$^{\circ}$ (i.e.: in the 
neutral line of the sunspot; see black region in the 
limb side penumbra of the middle panel of Fig.~1).
This is exactly what happens in our case. In Fig.~4 one
can see that although $\Delta v_{\rm los} \sim 3$ km s$^{-1}$ and
$\Delta \zeta \sim 40^{\circ}$ at all radial distances, the amount
of NCP produced (middle and bottom panels in Fig.~5) varies
with radial distance, reaching its maximum where $\tilde{\gamma} = 90-100^{\circ}$.

In addition to the average magnetic field inclination, the position of the upper and lower
boundaries of the flux tube $z^{*}=z_0 \pm R_{\rm t}$ play also a significant
role in producing NCP as we show below. Fig.~6 (upper panel) presents the position
(in the $\log\tau_5$ scale) of the upper (dashed) and lower (solid)
flux tube boundaries as a function of the radial distance in the penumbra.
The horizontal line at $\log\tau_5=0$ has been plotted in order to indicate when discontinuities
along the line of sight start producing a significant amount of
NCP in the Fe I lines at 6300 \AA~ (when $\tau_5 \leq 1$; see Fig.~8b in
Paper I). Our Fig.~6 shows that in the inner penumbra, the lower boundary of the flux tube
lies right beneath the continuum level, and therefore does not contribute
significantly to the generation of NCP. The area asymmetry
 observed at small radial distances (see Fig.~5; middle and bottom panels)
 is of the order of $\delta A \sim 0-20$ \% and is mainly due to the
upper flux tube boundary. At $r/R > 0.6$ the lower flux tube
boundary moves above the $\log\tau_5=0$ level and therefore starts contributing, 
together with the upper boundary, to the generation of area asymmetry. Note 
that it is at these radial distances that the observed $\delta A$ values are largest.

The bottom panel in Fig.~6 shows the radial variation of the 
penumbral flux tube diameter as inferred from the uncombed model.
According to our results the vertical extension of the penumbral 
flux tubes is roughly 100-300 km, in good agreement with
previous investigations based only on net circular polarization
considerations (Solanki \& Montavon 1993; Mart{\'\i}nez Pillet 2000,2001). 
It is also comparable to the horizontal width of the penumbral fibrils
observed in the continuum images at very high spatial resolution
by S\"utterlin (2001), Scharmer et al. (2002) and S\"utterlin et al. (2004).
Note that, although the penumbral flux tubes are thinner in the inner
penumbra on a geometrical scale, the thickness is almost constant with
radial distance in the optical depth scale. This opacity excess
is explained by the enhanced gas pressure and temperature 
inside the penumbral flux tubes at the inner penumbra. Note that
the vertical width of the flux tube is an upper limit in the sense that
a broader and tube would have difficulty to reproduce the observations.
The same is true to a narrower tube, if only a single tube lies along the line
of sight. However, we cannot rule out that multiple flux tubes lying on top
of each other cannot reproduce the observations.

Finally, in Fig.~7 we present a close up of the square region in Fig.~1
(middle panel) that has been analyzed in this work. The color code indicates
the observed area asymmetry (top panel) and the area asymmetry produced
by the uncombed model (bottom panel). The similarity is striking, with the
uncombed model being able to reproduce much (but not all) of the small scale 
behavior of the NCP.

\subsection{Magnetic flux conservation}%

As we have demonstrated the uncombed
model is able to reproduce with good accuracy the area asymmetry in
the circular polarization profiles using flux tubes of some 100-300 km in
diameter. In this section we check whether the model also satisfies the 
condition ${\bf \nabla} \cdot {\bf B}=0$. Let us consider cylindrical coordinates 
where the vectors ${\bf e_\rho}$ and ${\bf e_\phi}$ are contained in the 
plane perpendicular to the flux tube's axis ($\bar{YZ}$ plane in Fig.~2) 
and the vector ${\bf e_x}$ points along the tube's axis (i.e: radial direction 
along the penumbra). The null divergence condition is written as:

\begin{eqnarray}
{\bf \nabla} \cdot {\bf B} = \frac{1}{\rho}\frac{\partial}{\partial \rho}\left(\rho
  B_\rho\right)+\frac{1}{\rho}\frac{\partial B_{\phi}}{\partial \phi}
+\frac{\partial B_x}{\partial x} = 0
\end{eqnarray}

This condition must be satisfied locally by both the flux tube
and the surrounding atmosphere. Let us first focus on the surrounding
magnetic field ${\bf B_{\rm s}}$. The third term in Eq.~4 (derivative along the
tube's axis) can be written as:

\begin{eqnarray}
\frac{\partial B_{xs}}{\partial x} = \frac{\partial B_{rs}}{\partial r} = 
\frac{\partial }{\partial r}[B_{\rm s}(r) \sin \zeta_{\rm s}(r)]
\end{eqnarray}

\noindent where we considered that the $x$ coordinate (tube's axis) 
points radially in the penumbra. This in turn means that we can use
the the radial dependences of $B_{\rm s}$ and $\zeta_{\rm s}$
(zenith angle) from Fig.~4. It is clear that, the surrounding
magnetic field has to bend aside in order to accommodate the flux tube. 
This means that none of the first two terms in Eq.~4 will be
individually identical to zero. Under our model assumptions, these
two derivatives can not be computed. Nevertheless, assuming that
the null divergence condition is satisfied, we can calculate:

\begin{eqnarray}
-\frac{1}{\rho}\frac{\partial}{\partial \rho}\left(\rho B_{\rho s} \right)
- \frac{1}{\rho}\frac{\partial B_{\phi s}}{\partial \phi} =
\frac{\partial }{\partial r}[B_{\rm s}(r) \sin \zeta_{\rm s}(r)]
\end{eqnarray}

As far as the flux tube's magnetic field ${\bf B_{\rm t}}$ is concerned,
the magnetic field in our model is parallel to the tube's axis and therefore
$B_\rho$ and $B_\phi$ in Eq.~4 are identically zero. Interestingly,
a study of the magnetic configurations leading to a magnetohydrostatic
solution (Borrero 2004; Borrero et al. 2006, in preparation) shows that 
the flux tube's magnetic field must have a non vanishing component in the 
$\bar{YZ}$ plane. Since our model also samples the flux tube using
one ray path only, the first two terms in Eq.~4 can not be computed.
Once more, assuming that the null divergence condition is satisfied,
we can write:

\begin{eqnarray}
-\frac{1}{\rho}\frac{\partial}{\partial \rho}\left(\rho B_{\rho t} \right)
- \frac{1}{\rho}\frac{\partial B_{\phi t}}{\partial \phi} =
\frac{\partial }{\partial r}[B_{\rm t}(r) \sin \zeta_{\rm t}(r)]
\end{eqnarray}

\noindent where the right hand term can be deduced from the observations.
We have plotted in Fig.~8 the right hand terms of Eq.~6 and Eq.~7. 
This allows us to offer an upper limit on the derivatives along the $\bar{YZ}$ plane.
It turns out that both the flux tube and surrounding magnetic field 
are neraly divergence free in the plane perpendicular to the tube's axis.

\begin{eqnarray}
\frac{1}{\rho}\frac{\partial}{\partial \rho}\left(\rho B_{\rho} \right)
+ \frac{1}{\rho}\frac{\partial B_{\phi}}{\partial \phi} < 0.5 \;\;\; {\rm Gauss\; km^{-1}}
\end{eqnarray}

\begin{figure}
\begin{center}
\includegraphics[width=8cm]{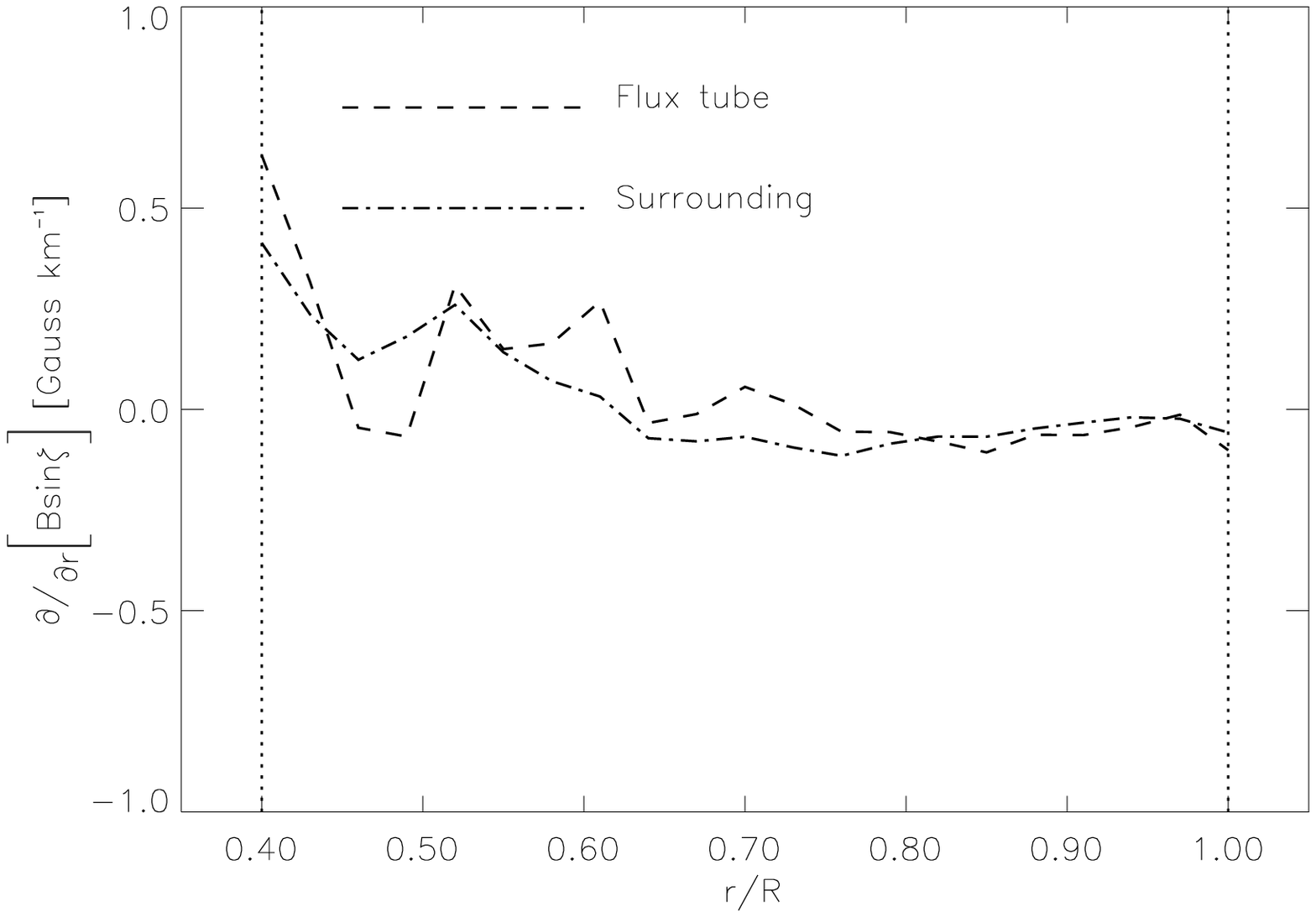}
\end{center}
\caption{$\frac{\partial }{\partial r}[B(r) \sin \zeta(r)]$ 
  as a function of the radial distance in the
  penumbra, for the external magnetic field (dot-dashed line) and the flux tube
  magnetic field (dashed line).}
\end{figure}

\section{Conclusions}%

The uncombed model is able to reproduce satisfactorily the polarization
signals emerging from the sunspot penumbra both for the Fe I lines at 6300
\AA~ and at 1.56 $\mu$m. The area asymmetry in the circular polarization
profiles, Stokes $V$, as well as its behavior with radial distance in the
penumbra are also reproduced. The azimuthal variation of the NCP observed in both sets
of spectral lines also finds a satisfactory explanation in terms of the uncombed
model (see M\"uller et al. 2002; Schlichenmaier et al. 2002).

We stress the important similarities of the results obtained in this work 
with the results from Papers I and II (compare with our Fig.~3 with Fig.~4 
in Paper I and Fig.~5 and 6 in Paper II). This is in spite of the fact that we 
have inverted different spectral lines (Fe I and OH lines at 1.56 $\mu$m
vs. Fe I \& Ti I lines at 630 nm). In addition, in our series of papers 
we have thoroughly studied three different sunspots at three 
different heliocentric angles, observed with two different telescopes. 
The remarkable consistency of the results provides strong support for the 
uncombed model. 

It also shows that the results obtained from the inversion of Stokes 
profiles are entirely consistent if interpreted in the context 
of this model (contrary to the claim of Spruit \& Scharmer 2005). 
Indeed, the ability to reproduce in detail a variety of
spectropolarimetric data must be the hallmark of any successful
model of the penumbral fine structure. 

Yet, some differences remain between the results obtained from the
Fe I 1.56 $\mu$m and Fe I 6300 \AA~ lines (see e.g.: the radial variation
of the flux tube filling factor in Fig.~4; cf. Fig.~5 in Paper II).
Recent developments in the existing instrumentation 
(see Socas-Navarro et al. 2005; Bellot Rubio \& Beck 2005)
now allow both spectral windows to be recorded simultaneously. 
It is planned to analyze such observations in order to resolve
these discrepancies. Such observations, if made at high resolution,
could allow for a simultaneous and accurate
determination of the vertical and horizontal structure of the sunspot penumbra.

It is important to note that the inferred vertical sizes for the penumbral flux
tubes (around 100-300 km) have been obtained from the interpretation of the
polarized spectrum, and in particular, of the net circular polarization, 
within the assumptions on the geometry of the penumbral fine structure
included in the uncombed field model. Although we can not rule out other models,
that offering a consistent explanation of the observed profiles, consider flux
tubes with smaller sizes, we are confident on the feasibility of the
uncombed model since it retrieves vertical extensions for the penumbral
fibrils of the same order of the horizontal widths observed by
Scharmer et al. (2002) and S\"utterlin et al. (2004) from continuum
images at high spatial resolution.

Flux tubes as thick as those retrieved here would imply that simulations of penumbral flux tubes
embedded in the penumbra (e.g Montesinos \& Thomas 1997, Schlichenmaier et al. 1998a,1998b)
would need to be revisited, since they all rely on the thin flux tube
approximation and therefore they neglect any variation in the physical
quantities in the plane perpendicular to the tube's axis. This will be the subject
of a future investigation: Borrero et al. (in preparation).

Last but not least we acknowledge that any model 
of the sunspot penumbra based on horizontal flux tubes embedded in a magnetic
surrounding atmosphere, has indeed a loose thread. As already pointed 
out by Schlichenmaier \& Solanki (2003) such models face serious difficulties to explain
the energy transport responsible for the penumbral brightness.

\begin{acknowledgements}
Computing time on a 198-CPU Linux cluster was generously provided by the 
"Gesellschaft f\"ur wissenschaftliche Datenverarbeitung G\"ottingen" (GWDG) and the 
"Max-Planck-Institut f\"ur Sonnensystemforschung" (MPS). Thanks to Rafael Manso-Sainz
for carefully reading the manuscript.
\end{acknowledgements}

\end{document}